\documentclass{JHEP3}
\usepackage{units,slashed}
\usepackage{amsmath}
\usepackage{epsfig}

\newcommand{\be}{\begin{equation}}
\newcommand{\ee}{\end{equation}}
\newcommand{\ba}{\begin{eqnarray}}
\newcommand{\ea}{\end{eqnarray}}
\newcommand{\MSb}{$\overline{\text{MS}}$}
\newcommand{\lsim}{\lesssim}
\newcommand{\gsim}{\gtrsim}
\newcommand{\tr}{\mathop\mathrm{tr}}
\newcommand{\cL}{\mathcal{L}}
\newcommand{\cU}{\mathcal{U}}
\newcommand{\e}{\mathrm{e}}

\title{Standard Model Higgs boson mass from inflation: two loop
  analysis}

\author{F. Bezrukov\\
  Max-Planck-Institut f\"ur Kernphysik,\\
  PO Box 103980, 69029 Heidelberg, Germany;\\
  Institute for Nuclear Research of the Russian Academy of Sciences,\\
  60th October Anniversary prospect 7a, Moscow 117312, Russia\\
  E-mail: \email{Fedor.Bezrukov@mpi-hd.mpg.de}}

\author{M. Shaposhnikov\\
  Institut de Th\'eorie des Ph\'enom\`enes Physiques,\\
  \'Ecole Polytechnique F\'ed\'erale de Lausanne,\\
  CH-1015 Lausanne, Switzerland\\
  E-mail: \email{Mikhail.Shaposhnikov@epfl.ch}
}

\abstract{We extend the analysis of \cite{Bezrukov:2008ej} of the
  Standard Model Higgs inflation accounting for two-loop radiative
  corrections to the effective potential.  As was expected, higher
  loop effects result in some modification of the interval for allowed
  Higgs masses $m_\mathrm{min}< m_H < m_\mathrm{max}$, which somewhat
  exceeds the region in which the Standard Model can be considered as
  a viable effective field theory all the way up to the Planck scale.
  The dependence of the index $n_s$ of scalar perturbations on the
  Higgs mass is computed in two different renormalization procedures,
  associated with the Einstein (I) and Jordan (II) frames.  In the
  procedure I the predictions of the spectral index of scalar
  fluctuations and of the tensor-to-scalar ratio practically do not
  depend on the Higgs mass within the admitted region and are equal to
  $n_s=0.97$ and $r=0.0034$ respectively.  In the procedure II the
  index $n_s$ acquires the visible dependence on the Higgs mass and
  and goes out of the admitted interval at $m_H$ below
  $m_\mathrm{min}$.  We compare our findings with the results of
  \cite{DeSimone:2008ei}.}


\date{\today}

\begin{document}

\section{Introduction}
\label{sec:intro}
A theory of inflation provides an attractive explanation of the basic
properties of the Universe, including its flatness, homogeneity and
isotropy (for a recent review see \cite{Linde:2007fr}). In addition,
it gives a natural source for generation of an almost flat spectrum of
fluctuations, leading to the structure formation (for a comprehensive
discussion see \cite{Mukhanov:2005sc}). Many realisations of
inflationary scenario introduce new physics between the electroweak
and Planck scales, and postulate the existence of a special scalar
field~--- inflaton. Different models of inflation lead to distinct
predictions of the main inflationary parameters, such as the spectral
index $n_s$ of the scalar perturbations and the tensor-to-scalar ratio
$r$. A number of models have been already excluded or are in tension
with the cosmological observations of the cosmic microwave background
(CMB) \cite{Komatsu:2008hk}.

Within the variety of inflationary models there is one which plays a
special role. It does not require introduction of any new physics and
identifies the inflaton with the Higgs field of the Standard Model
(SM) \cite{Bezrukov:2007ep}. The key observation which allows such a
relation is associated with a possible non-minimal coupling of the
Higgs field $H$ to the gravitational Ricci scalar $R$,
\be
  \label{nonmin}
  \cL_\mathrm{non-minimal} = \xi H^\dagger H R
  \;.
\ee
For large Higgs backgrounds $\xi h^2  \gsim M_P^2$ (here
$M_P=2.4\times \unit[10^{18}]{GeV}$ is the Planck scale and $h^2 = 2
H^\dagger H$) the masses of all the SM particles \emph{and} the induced
Planck mass $[M^\mathrm{eff}_P]^2 = M_P^2 + \xi h^2$ are proportional to
one and the same parameter, leading to independence of physical
effects on the magnitude of $h$. In other words, the Higgs potential
in the large-field region is effectively flat and can result in
successful inflation. The fact that the non-minimal coupling of the
inflaton to Ricci scalar relaxes the requirement for the smallness of
the quartic scalar self-interaction was noted already in
\cite{Spokoiny:1984bd,Salopek:1988qh,Fakir1990,Komatsu:1999mt,
Tsujikawa:2004my,Barvinsky1994,Barvinsky1998}.

This model of inflation is very conservative in a sense that it does
not require the existence of new particles and interactions: the term
(\ref{nonmin}) is needed for the renormalizability of the SM in the
curved space-time. Since only \emph{one} new parameter is introduced
(the value of the non-minimal coupling $\xi$, fixed by the amplitude
of scalar perturbations at the COBE scale \cite{Bezrukov:2007ep}) the
Higgs inflation is a predictive theory, allowing to fix a number of
parameters that can be found by the observations of the CMB.
Moreover, the model does not have a problem of the graceful exit from
the inflationary regime and leads to high values of the reheating
temperature, $T_\mathrm{r} \sim \unit[10^{13}]{GeV}$
\cite{Bezrukov:2008ut}.  The value of $\xi$, necessary for successful
inflation, is required to be rather large, $\xi\sim 10^3-10^4$.

Due to the minimal character of the Higgs inflation it can be ruled
out by the CMB observations \emph{and} by particle physics
experiments, especially those at the LHC.  In
refs.~\cite{Bezrukov:2008cq,Shaposhnikov:2008rc} we conjectured that
for the Higgs boson to play a role of the inflaton the SM has to be a
consistent field theory all the way up to the Planck scale. This puts
the mass of the SM Higgs to a specific interval. The mass is
constrained from above by the requirement that no Landau pole appears
in the scalar self-coupling of the Higgs boson below the Planck scale
(see \cite{Maiani:1977cg,Cabibbo:1979ay,Lindner:1985uk,Hambye:1996wb}
and references therein). The lower limit comes from the demand that
the electroweak vacuum is stable for all scalar field values below
$M_P$ (see
\cite{Krasnikov:1978pu,Hung:1979dn,Politzer:1978ic,Altarelli:1994rb,Casas:1994qy,Casas:1996aq}
and references therein). Our results were challenged in
\cite{Barvinsky:2008ia}, claiming that radiative corrections to the
tree Higgs potential are large in the inflationary region and that the
spectral index $n_s$ of scalar fluctuations is only consistent with
WMAP observations for values of the Higgs mass
$m_H\gtrsim\unit[230]{GeV}$.  The latter conclusions were shown to be
modified drastically by the running of the coupling constants in
\cite{Bezrukov:2008ej} and \cite{DeSimone:2008ei}, containing the
one-loop and two-loop renormalization group (RG) improved analysis
respectively. A similar remark was also made in
\cite{GarciaBellido:2008ab}.

The aim of the present work is to upgrade the computation of
\cite{Bezrukov:2008ej} to the two-loop level.  This allows to check
the stability of our conclusions against radiative corrections and to
make a proper comparison with \cite{DeSimone:2008ei}, where two-loop
computation was presented.

The paper is organised as follows. In section \ref{sec:assumptions} we
discuss the basic assumptions and elaborate on the possible choices of
normalization point for radiative corrections.  In section
\ref{sec:action} we consider the action of the Standard Model in the
inflationary region, in section \ref{sec:potential} we compute the
two-loop effective potential in the inflationary region. In section
\ref{sec:rg} we discuss the RG improvement of the effective potential.
In section \ref{sec:procedure} we fix the procedure for computing the
radiative corrections to the inflationary potential. In section
\ref{sec:num} we present the numerical results. In section
\ref{sec:comp} we discuss the difference between our analysis and that
of ref.~\cite{DeSimone:2008ei}. Section \ref{sec:conc} is conclusions.
We use the same notations as in \cite{Bezrukov:2008ej}. The present
work can be considered as a continuation of this paper. In particular,
we omit all the discussions already made in \cite{Bezrukov:2008ej}.

\section{The basic assumptions}
\label{sec:assumptions}
It is a challenge to invent a self-consistent framework for
computation of radiative corrections to inflation in any type of field
theory. The reason is associated with non-renormalizable character of
gravity.  To clarify this point, let us consider the most popular  line of reasoning which is based on the following logic
(see, e.g.\ \cite{Burgess:2009ea,Barbon:2009ya}), related to
construction of effective field theories.

Take some field theory coupled to gravity and consider cross-sections
of different reactions (such as scalar-scalar or graviton-graviton
scattering) computed in the lowest order of perturbation theory. Find
the lowest energy $\Lambda$ at which one of these cross-sections hits
the unitarity bound. Call $\Lambda$ the ``ultraviolet cutoff''. The
fact that perturbation theory breaks down for energies above $\Lambda$
may then signal that the theory under consideration is not a
fundamental theory, but an effective one, valid only at momenta
smaller than the ultraviolet cutoff $\Lambda$.\footnote{This type of
arguments were very successful for construction of renormalizable
gauge theories with spontaneous symmetry breaking
\cite{Cornwall:1974km} in general and the SM in particular, starting
from four-fermionic Fermi interaction.}  If true, the initial
Lagrangian \emph{must} be modified by adding to it all sorts of
higher-dimensional operators, suppressed by the scale $\Lambda$. Now,
if the addition of these higher dimensional operators with
coefficients of the order of one spoils the effect under
consideration, it is said that the realisation of a particular
phenomenon (inflation in our case) is ``unnatural''. At the same time,
if one finds some symmetry principle which keeps the unwanted
contributions of higher dimensional operators small or zero, then it
is said that the corresponding effect is ``natural''.\footnote{Of
course, the real physics question is not whether this or that theory
is ``natural'' but whether it is realised in Nature. The small value
of the scale of the electroweak symmetry breaking in comparison with
the Planck scale is considered to be extremely ``unnatural''. The same
is true for the cosmological constant. Nevertheless, these are the
values we have to live with.}

For the values of the non-minimal couplings $\xi \gg 1$ the ``cutoff''
scale, determined by the method described above, is of the order
of:\footnote{This estimate was done by S. Sibiryakov (private
communication) and later in \cite{Burgess:2009ea,Barbon:2009ya}.}
\be
\Lambda_\xi \sim \frac{M_P}{\xi} \ll M_P\;.
\label{cutoff}
\ee
If the higher order operators, suppressed by $\Lambda_\xi$, are added
to the SM with coefficients of the order of one, they spoil the
Higgs-inflation, and, according to the definition discussed above,
make it ``unnatural'' \cite{Burgess:2009ea,Barbon:2009ya}. In other
words, the inflationary predictions depend in a sensitive way on the
coefficients with which these operators appear.

The point of view we take in this paper is very much different from
the reasoning described above. Contrary to
\cite{Burgess:2009ea,Barbon:2009ya}, we will \emph{assume} that the
breaking of perturbation theory, constructed near the SM vacuum at
energies greater than $\Lambda$, is not a signal of existence of new
physics which should replace the SM, but rather an indication of a new
physical phenomenon~--- strong coupling, which should be treated
\emph{inside} the SM by non-perturbative methods (such as resummation
of different diagrams, lattice simulations, etc.). Put it in other
words, we will assume that the SM is valid for all momenta smaller
that the Planck mass $M_P$. Therefore, no higher dimensional
operators, suppressed by $\Lambda_\xi$ will be added\footnote{As has
  been shown in \cite{Bezrukov:2008ut}, the higher order operators
  suppressed by $M_P$ are harmless for inflation.}.

Fortunately, the presence of the strong coupling at energies higher
than $\Lambda_\xi$ in scattering of particles defined as excitations
above the \emph{SM ground state} (with small vacuum expectation value
of the Higgs field) does not prevent the reliable computation of the
inflationary potential. As we will see (section \ref{sec:action}), in
the inflationary region, for the Higgs field values $h_\mathrm{inf}
\sim M_P/\sqrt{\xi}$, an adequate description of the system is given
by the so-called chiral electroweak theory \cite{Longhitano:1980iz}
(SM with the frozen radial Higgs mode). This theory is not
renormalizable and has an ultraviolet cutoff
$\Lambda_\mathrm{inf}\simeq h_\mathrm{inf}$, defined with the use of
the procedure, discussed above. $\Lambda_\mathrm{inf}$ is associated
with the scattering of the longitudinal intermediate vector
bosons.\footnote{Not surprisingly, the energy at which the tree
  cross-sections start to exceed the unitarity bound, depends on the
  background field. For example, for the graviton-graviton scattering
  the corresponding cutoff is $[M^\mathrm{eff}_P]^2$, rather than
  $M_P^2$.}  At the same time, the typical energy saturating the loop
diagrams appearing in the computation of the effective potential is of
the order of the masses of the SM particles in the Higgs background
(dimensional regularisation is assumed) and thus is smaller than the
cutoff by the magnitude of gauge or Yukawa coupling constants. In
other words, the contribution of the strongly interacting domain of
energies to the low-energy observables such as the effective potential
is suppressed.

In spite of the fact that we do not include higher-order operators to
the computation, the effective potential \emph{cannot} be fixed
unambiguously. In section~3 of \cite{Bezrukov:2008ej} we discussed two
distinct possibilities. In the first one (referred as prescription I)
the normalization point $\mu$ (coinciding, for example,  with
t'Hooft-Veltman parameter in dimensional regularisation) of the loop
integrals is chosen to be proportional to the coefficient in front of
the scalar curvature. In the Jordan frame this gives
\be
\mu^2 \propto M_P^2 +\xi h^2\;,
\ee
and, consistently, in the Einstein frame
\be
\mu^2 \propto M_P^2 \;.
\label{einI}
\ee
With these choices the computations in Jordan and Einstein frames are
equivalent both at classical and quantum levels. The prescription II
corresponds to the Jordan frame normalization point given by
(\ref{einI}). For consistency, one has to choose then the
normalization point in the Einstein frame as
\be
\mu^2 \propto M_P^4/(M_P^2 +\xi h^2) \;.
\label{einII}
\ee
In short, the prescription I is ``standard'' (field-independent)
in the Einstein frame, whereas the prescription II is ``standard'' in
the Jordan frame.

A possible physics behind the choice II is related to an
idea that the Jordan frame is the one in which ``distances are
measured'' \cite{Barvinsky:2008ia} and that in this frame there exist
a fundamental cut-off, independent on background scalar field. A
possible physics leading to the choice I is related to quantum scale
invariance, discussed in \cite{Shaposhnikov:2008xi}, though the
prescription II can be easily realised in scale-invariant theories as
well.\footnote{In the language of \cite{Shaposhnikov:2008xi} the
  prescription I corresponds to the choice of the dimensional
  regularisation parameter $\mu^2\propto\xi_\chi\chi^2+\xi_hh^2$, while
  prescription II corresponds to the choice $\mu^2\propto\chi^2$,
  where $\chi$ is the dilaton field.}

To elucidate the difference between these two prescriptions let us
consider the limit of large Higgs fields, $\xi h^2 \gg M_P^2$. In this
case the dimensionful parameters $M_P^2$ and the Higgs mass can be
neglected, and the Lagrangian of the electroweak theory with gravity
incorporated acquires a new symmetry (at the classical level)~--- the
scale invariance, $\phi(x)\to \lambda^\alpha\phi(\lambda x)$, where
$\phi$ is a generic notation for any field with canonical mass
dimension ${\rm GeV}^\alpha$. Note that in the presence of gravity the
scale invariance does not ensure conformal invariance, since $\xi \neq
- \frac{1}{6}$. The fate of the dilatation symmetry at the quantum
level depends on the way the infinities arising in loop computations
are regularized and subtracted.

The use of standard regularization and renormalization schemes (such
as dimensional or Pauli-Villars) leads to the trace anomaly in
energy-momentum tensor and thus to the breaking of dilatation symmetry
(for a review see \cite{coleman}). The flat direction, existing for
the Higgs field in Einstein frame, gets bended owing to radiative
corrections. This is the prescription II. If gravity is taken away, it
leads to a \emph{renormalizable} field theory.

A non-conventional approach is to use dimensional regularization
($n=4-2\epsilon$) and keep the scale-invariance intact even if
$\epsilon \neq 0$ \cite{Englert:1976ep,Shaposhnikov:2008xi} (see also
\cite{Jain:2008qv}) by replacing the t'Hooft-Veltman parameter $\mu$
by a combination of dynamical fields with matching mass dimension.  In
this case the theory stays scale-invariant at the quantum level  and
the flat direction remains intact. This is the prescription I. If
gravity is taken away, it leads to an \emph{effective} field theory,
valid for momenta transfers up to the effective Planck scale $\sim
\sqrt{\xi} h$ \cite{Shaposhnikov:2009nk}.

It is the trace (or conformal) anomaly in the total matter
energy-momentum tensor  which leads to the dependence of the
inflationary predictions on the Higgs mass.  The one- and and two-loop
corrections to the effective potential, that we find in this paper in
the inflationary region, account exactly for this anomaly, which is
different in prescriptions I and II.

It looks impossible to fix the unique prescription without knowing the
physics at the Planck scales. In other words, the determination of
inflationary cosmological parameters, such as the spectral index $n_s$
through parameters of the SM, is subject to an intrinsic uncertainty,
related to renormalization procedure. We will do computations with
both prescriptions. Fortunately, the numerical difference happens to
be small in two-loop approximation, as it was so at one-loop
\cite{Bezrukov:2008ej}.

\section{Action in the inflationary region}
\label{sec:action}
After transformation to the Einstein frame the electroweak Lagrangian is
essentially non-linear and is given by
\begin{equation}
  \label{chiral}
  \cL_\mathrm{chiral} =
  \frac{1}{2}(\partial_\mu \chi)^2-\frac{1}{g^2}H_1
  -\frac{1}{g'^2}H_2- L_{W/Z} + L_Y - U(\chi)
  \;,
\end{equation}
where
\begin{align}
  H_1 &= \frac{1}{2}\tr[W_{\mu\nu}^2]\;,\\
  H_2 &= \frac{1}{4}B_{\mu\nu}^2\;,\\
  L_{W/Z} &= \frac{v^2}{4}\tr[V_\mu^2]\;,\\
  L_Y &=i\bar Q_{L,R}\slashed D Q_{L,R} -(\frac{y_t v}{\sqrt{2}} \bar Q_L \tilde \cU Q_R + \dots + \mathrm{h.c.})
  \;,\label{yukawa}
\end{align}
and $D$ is the standard covariant derivative for the fermions.
The field $\chi$ is related to the module of the Higgs field $h$
through the equation
\begin{equation}
  \frac{d\chi}{dh}=\sqrt{\frac{\Omega^2 + 6\xi^2h^2/M_P^2}{\Omega^4}}
  \;,
\end{equation}
with the conformal factor $\Omega(h)$ given by
\be
\Omega^2(h)=  1+ \frac{\xi h^2}{M_P^2}\;.
\ee
For $h \ll M_P/\sqrt{\xi}$ the relation between $\chi$ and $h$ is
\begin{equation}
  \chi \simeq \frac{h}{2}\left[
    \sqrt{1+6\xi^2h^2/M_P^2}
    +\frac{\mathop\mathrm{arcsinh}\left(\sqrt{6}\xi
        h/M_P\right)}{\sqrt{6}\xi h/M_P}
  \right]
  \;,
\end{equation}
and for $h \gg M_P/\xi$ we have
\be
\chi \simeq \sqrt{\frac{3}{2}}M_P\log\Omega^2(h)\;.
\ee
The parameter $v$ in the inflationary case is
\be
v^2(h) = \frac{h^2}{\Omega^2(h)}\;.
\ee
The dimensionless Nambu-Goldstone bosons $\pi^a$
are parametrized in the non-linear form as
\begin{equation}
  \cU= \exp\left[2i\pi^a T^a\right]
  \,,\quad
  T^a=\frac{\tau^a}{2}
  \;.
\end{equation}
In the Yukawa Lagrangian we only write explicitly the top quark
contribution, and define the quark fields as
\be
  \bar Q_L =(\bar t_L,\bar b_L)\,,\;
  \tilde \cU = -\tau_2 \cU^*\tau_2\,,\;
  \bar Q_R =(\bar t_R, 0)\;,
\ee
where $\tau_a$ are the Pauli matrices.

The gauge fields and the fields strength are given by
\begin{align}
  V_\mu &= (\partial_\mu \cU) \cU^\dagger + i W_\mu -i \cU B_\mu^Y \cU^\dagger
  \;,\\
  W_\mu &= W_\mu^a T^a\;, & B_\mu^Y &= B_\mu T^3
  \;,\\
  W_{\mu\nu} &= \partial_\mu W_\nu - \partial_\nu W_\mu +i[W_\mu,W_\nu]
  \;, & B_{\mu\nu} &= \partial_\mu B_\nu - \partial_\nu B_\mu
  \;.
\end{align}
The scalar potential is
\begin{equation}
  \label{U}
  U(\chi) = \frac{\lambda h^4(\chi)}{4\Omega^4}
  \;,
\end{equation}
where we have neglected the Higgs mass, irrelevant for analysis of
inflation. For $h \gg M_P/\xi$ we have
\begin{equation}
  \label{U(chi)}
  U(\chi) \simeq \frac{\lambda M_P^4}{4\xi^2}
  \left(
    1-\e^{-\frac{2\chi}{\sqrt{6}M_P}}
  \right)^{2}
  \;.
\end{equation}

As has been discussed in \cite{Bezrukov:2008ej}, in the background of
the ``small'' Higgs fields $h\lsim\frac{M_P}{\xi}$, the action in the
Einstein frame coincides with that of the canonical Standard Model.
Thus, in this region the computation of the effective potential is
straightforward (and in fact not necessary, as the inflation takes
place at higher values of the scalar field).

In the inflationary region the Higgs field is of the order
\begin{equation}
  \label{infreg}
  h\gtrsim\frac{M_P}{\sqrt{\xi}}
  \;,
\end{equation}
corresponding to $\chi\gtrsim M_P$.  For $\xi \gg 1$ the Higgs field is
essentially massless in the region (\ref{infreg}), and decouples from
all the fields of the SM.  The transition between these two regimes is
governed by the effective couplings of the deviation $\delta \chi$ of
the field $\chi$ from its background value $\chi_c$ to the fields of
the SM.  The canonical values of the Higgs field couplings to other
fields are multiplied by the factor $q$ given by
\be
q=\frac{d[h/\Omega]}{d\chi}=\frac{1}{\Omega}\frac{1}{\sqrt{\Omega^2 +
    6\xi^2h^2/M_P^2}}\;.
\label{qfactor}
\ee
For small fields $h\ll M_P/\xi$ the factor $q$ is equal to the SM value
$q = 1$ and for large fields $M_P/\xi \ll h \lsim M_P/\sqrt{\xi}$ it
leads to the  suppression of Higgs interactions by  $q =
M_P/(\sqrt{6}\xi h)\ll 1$. This effect can also be seen in the Jordan
frame \cite{Barvinsky:2008ia,DeSimone:2008ei}, where it emerges due to the
kinetic mixing between the Higgs field and gravitational degrees of
freedom. The relation between the suppression factor $s$ found in
\cite{DeSimone:2008ei} and the factor $q$ is $s=q^2\Omega^4$.

In the limit $h\to\infty$ the action (\ref{chiral}) is nothing but the
chiral electroweak theory studied in detail in numerous papers (for a
review see, e.g.\ \cite{Feruglio:1992wf}) plus a massless
non-interacting field.  It will be used for analysis of radiative
corrections to the inflationary potential below.

\section{Two-loop effective potential}
\label{sec:potential}
The effective potential for the field $\chi$ is gauge-invariant, since
$\chi$ is gauge-invariant itself. However, the computation is most
convenient in Landau gauge, in which the Goldstone bosons (hidden in
the matrix $\cU$) and ghosts are massless, and a number of diagrams can
be dropped due to the fact that the vector propagators are transverse
\cite{Coleman:1973jx}. The one-loop contribution to the potential is
given by
\begin{equation}
  \label{U1loop}
  U_1 =
  \frac{6m_W^4}{64\pi^2}\left(\log\frac{m_W^2}{\mu^2}-\frac{5}{6}\right)+
  \frac{3m_Z^4}{64\pi^2}\left(\log\frac{m_Z^2}{\mu^2}-\frac{5}{6}\right)
  - \frac{3m_t^4}{16\pi^2}\left(\log\frac{m_t^2}{\mu^2}-\frac{3}{2}\right)
  \;,
\end{equation}
where we keep now the constant terms omitted in Eq.~(9) of
\cite{Bezrukov:2008ej}, as they are needed for a consistent two-loop
analysis.  As in \cite{Bezrukov:2008ej}, the part associated with the
(light in the inflationary region) field $\chi$ is neglected.  The
background field masses of intermediate bosons ($m_W$ and $m_Z$), and
of t-quark ($m_t$) are given in Eq.~(8) of \cite{Bezrukov:2008ej}.

To find the potential in two loops, one should first expand $\cU$ with
respect to Goldstone fields $\pi^a$. It is sufficient to keep first
order terms in the Yukawa part (\ref{yukawa}) and second order terms in
the gauge part (\ref{chiral}). After this is done, the computation of
the two-loop contribution is straightforward but rather involved.
Fortunately, all necessary contributions can be extracted from
\cite{Ford1992}. This paper found the following representation of
the two-loop contribution:
\be
  U_2 = V_S+V_{SF}+V_{SV}+V_{FV}+V_V\;,
  \label{egeneral}
\ee
where $V_S$ represents scalar loops, $V_{SF}$---the diagrams involving
scalars and fermions, $V_{SV}$---scalars and vectors,
$V_{FV}$---fermions and vectors, and $V_V$---vectors and ghosts. To
get the effective potential for our case one has to take away all
diagrams involving the Higgs field, since the interactions of it with
itself and all matter fields are suppressed by $\xi$ in the
inflationary region. In addition, all Goldstone masses have to be put
to zero. As a result, the following substitutions have to be made:
\begin{align}
  \label{repl}
  V_{S} &\to 0 \;
  ,\\
  V_{SF} &\to
  m_t^2\left(m_t^2\hat I(m_t,0,0)+\frac{1}{2}\hat J(m_t,m_t)\right)
  \;,\\
  V_{SV} &\to \frac{1}{4}g^2\left[
    \frac{(1-2\sin^2\theta)^2}{2\cos^2\theta}A(0,0,m_Z)
    +  A(0,0,m_W)\right]
  \nonumber\\
  &\phantom{\to} - g^2 \sin^4\theta m_Z^2 B(m_Z,m_W,0) -
  e^2 m_W^2 B(m_W,0,0)
  \;,
\end{align}
whereas $V_{FV}$ and $V_V$ stay intact. Explicit expressions for
different functions appearing here are quite long and we do not
reproduce them, see \cite{Ford1992} for all definitions.

The two-loop effective potential derived in this way contains an
explicit dependence on the normalization point $\mu$. It is, however,
spurious as it must be compensated by the running of the coupling
constants and the field $\chi$.  If we knew the RG running of the
parameters, the best choice of $\mu$ would be the background mass of a
vector boson or of t-quark, which will minimize the logarithms in the
higher-loop corrections.  In the next section we will derive the
necessary RG equations, valid for large field values in the
inflationary region.

\section{Renormalization group equations}
\label{sec:rg}
The change of the scalar field from the border-line of applicability
of the canonical Standard Model $h \sim M_P/\xi$  to the inflationary
region  $h_\mathrm{inf}\sim M_P/\sqrt{\xi}$ is relatively small. This
means that the \emph{one-loop} renormalization improvement of the
\emph{two-loop} effective potential will do already a very good job.
The
main objects which appear in the one-loop effective potential
(\ref{U1loop}) are the background masses of $W$, $Z$ bosons and of $t$-quark.
Therefore, we need to find the logarithmic running of these masses. It
is sufficient to consider their renormalization in the asymptotic
region of large $h$, where the tree potential for the scalar field is
flat. Here the action (\ref{chiral}) is just the action of the chiral
SM with $v=M_P/\sqrt{\xi}$.

To find the running of $m_Z$, $m_W$ and $m_t$ it is sufficient to
consider the $WW$, $ZZ$ and $tt$ two-point functions in the chiral SM.
The explicit computation gives (in the \MSb{} subtraction scheme):
\begin{align}
  \label{mw}
  16\pi^2 \mu\frac{\partial}{\partial\mu} m_W^2
  &=
  \left(\frac{3g'^2}{2}-\frac{23g^2}{2}-6y_t^2+\frac{8g^2
      n_f}{3}\right)m_W^2
  \;,\\
  \label{mz}
  16\pi^2 \mu\frac{\partial}{\partial\mu} m_Z^2
  &=\left(
    \frac{40n_f g'^4}{9}+\frac{g'^4}{6}+\frac{8g^2 g'^2}{3}-6 y_t^2 g'^2-
    \frac{23g^4}{2}-6 g^2 y_t^2+\frac{8 g^4n_f}{3}\right)\frac{v^2}{4}
  \;,\\
  \label{mt}
  16\pi^2 \mu\frac{\partial}{\partial\mu} m_t^2
  &=-\left(
    \frac{4 g'^2}{3}+16 g_3^2\right)m_t^2
  \;,
\end{align}
where $n_f=3$ is the number of fermionic generations.

In fact, it is more convenient to rewrite these equations by
introducing the running of the coupling constants $g'$, $g$,
$g_3$, $y_t$, $\lambda$, and $\xi$.  As the chiral SM is well studied in
the literature, we can extract the necessary results from the existing
computations. Reference~\cite{Dutta:2007st} gives the following equations
for the running of $g'$, $g$:
\begin{align}
  \label{g'}
  16\pi^2 \mu\frac{\partial}{\partial\mu}g'
  &=
  \left(\frac{1}{6}-\frac{1}{12}+\frac{20n_f}{9}\right)g'^3
  \;,\\
  \label{g}
  16\pi^2 \mu\frac{\partial}{\partial\mu}g
  &=
  -\left(\frac{43}{6}+\frac{1}{12}-\frac{4n_f}{3}\right)g^3
  \;.
\end{align}
The number $\frac{1}{12}$ accounts for the fact that the Higgs field
(but not Nambu-Goldstone bosons) decouples from the fields of the SM.

Reference~\cite{Dutta:2007st} also gives an equation for running of the
parameter $v^2$, without accounting for the Yukawa couplings. It is
not difficult to add the corresponding contribution, what leads to:
\begin{equation}
  \label{v}
  16\pi^2 \mu\frac{\partial}{\partial\mu}v^2=
  \left(\frac{3}{2}g'^2+3g^2-6y_t^2\right)v^2
  \;.
\end{equation}
Since $v^2 \propto \frac{1}{\xi}$ in our case, this can be converted
to equation for $\xi$:
\be
16\pi^2 \mu\frac{\partial}{\partial\mu}\xi=
-\left(\frac{3}{2}g'^2+3g^2-6y_t^2\right)\xi\;.
\label{xi}
\ee
Clearly, the running of the strong coupling does not change in the
chiral phase, so that
\begin{equation}
  \label{g3}
  16\pi^2 \mu\frac{\partial}{\partial\mu}g_3 = -7 g_3^2
  \;.
\end{equation}
Now, combination of Eq.~(\ref{mt}) with (\ref{v}) gives an equation for the
top Yukawa coupling
\be
16\pi^2 \mu\frac{\partial}{\partial\mu}y_t=
\left(-\frac{17}{12}g'^2 - \frac{3}{2}g^2 - 8g_3^2 + 3
y_t^2\right)y_t\;.
\label{yt}
\ee The equation for $\lambda$ follows from equations for $v^2$ and
from the fact that the one-loop effective potential
$U(h,\lambda(\mu))+U_1(h,\mu)$
in the asymptotic region of the Higgs
fields must be independent on $\mu$:
\begin{equation}
  \label{lambda}
  16\pi^2\mu\frac{\partial}{\partial\mu}\lambda =
  -6y_t^4+\frac{3}{8} \left( 2g^2+(g'^2+g^2)^2 \right)
  + \left( -3g'^2 - 6 g^2 + 12y_t^2 \right) \lambda
  \;.
\end{equation}
This equation is quite peculiar as it does not contain the term
$\lambda^2$ and thus does not have the Landau-pole behaviour at one
loop.  It is instructive to provide also the equation for the
combination $\lambda/\xi^2$, which is the only combination of
constants present in the tree level potential (\ref{U(chi)})
\begin{equation}
  \label{eq:lambdaxi}
  16\pi^2\mu\frac{\partial}{\partial\mu}\left(\frac{\lambda}{\xi^2}\right)
  =
  \frac{1}{\xi^2}\left(
    -6y_t^4+\frac{3}{8} \left( 2g^2+(g'^2+g^2)^2 \right)
  \right)
  \;.
\end{equation}
This equation does not contain terms proportional to $\lambda$ at all,
making the running less important for heavier Higgs masses. It is
interesting to note that the right-hand-side of Eq.
(\ref{eq:lambdaxi}) is exactly the same for $h \lesssim M_P/\xi$, what
can be derived from eqns. (10-14) of \cite{Bezrukov:2008ej}.

However, this system of RG equations is not complete. The reason is
that the chiral SM is not renormalizable, so that in order to remove
the divergences one has to add a certain set of counter-terms with the
structures different from those already present in (\ref{chiral}).
These operators were found and discussed in detail in a number of
papers (see, e.g.\ a recent work \cite{Dutta:2007st}). Only few of
these operators are needed for our purposes, namely those which
contribute to the renormalization of the mass of $Z$
boson.\footnote{Note that the renormalization of $W$ and $t$ masses
  does not require any new counter-terms, at least at the order in the
  coupling constants we are working with.} These are:\footnote{A
  potentially important operator corresponding to $\alpha_8$ (see
  \cite{Dutta:2007st} for definition) does not contribute at the order
  in the coupling constants we are interested in.}
\be
\nonumber
\frac{1}{4} \alpha_0 v^2\left(\tr[\Upsilon V_\mu]\right)^2+
\frac{1}{2} \alpha_1 B_{\mu\nu}\tr[\Upsilon W^{\mu\nu}]\;,
\ee
where $\Upsilon = 2 \cU T^3 \cU^\dagger$. The coefficients  $\alpha_0$ and
$\alpha_1$ obey the RG equations \cite{Dutta:2007st}
\be
16\pi^2 \mu\frac{\partial}{\partial\mu}\alpha_0 = \frac{3}{4} g'^2
\;,\qquad
16\pi^2 \mu\frac{\partial}{\partial\mu}\alpha_1 = \frac{1}{6}
\label{alpha}
\ee
and the mass of the $Z$-boson is
\be
m_Z^2 =
\frac{1}{4}\left((g'^2+g^2)(1-2\alpha_0)-2g'^2g^2\alpha_1\right)v^2\;.
\label{Zmass}
\ee
This expression, supplemented by the RG equations
(\ref{g'}), (\ref{g}), (\ref{g3}), (\ref{xi}), (\ref{yt}),
(\ref{lambda}), (\ref{alpha}),
allows for a one-loop improvement of the 2-loop effective potential in
the inflationary region. The initial conditions for the running are to
be fixed at the borderline between the SM phase and the SM chiral
phase, corresponding to $\mu \simeq M_P/\xi$.

\section{The procedure for computations of inflationary parameters}
\label{sec:procedure}

To determine the inflationary parameters, we generalize the procedure
described in \cite{Bezrukov:2008ej} to two loop level. It consists of
several steps described below.

\begin{enumerate}
\item To find the initial conditions for the RG running of the
  coupling constants at $\mu=m_t$ we relate the values of the
  \MSb-scheme $g$, $\lambda$ and $y_t$ to the pole Higgs
  and top masses with the use of the formulas given in section 5 of
  ref.~\cite{Kajantie:1995dw} and in the appendix of
  ref.~\cite{Espinosa:2007qp}. The procedures described in these works
  give almost identical results (see appendix).

\item We replace the system of one-loop RG equations (10-14) of
  \cite{Bezrukov:2008ej} for all three gauge couplings, scalar
  self-coupling and top-quark Yukawa couplings by the corresponding
  two-loop equations taken from \cite{Luo:2002ey} and reproduced in
  convenient form in \cite{Espinosa:2007qp}. Solving these equations up
  to the scale $M_P/\xi$ gives us the values of parameters at the
  borderline between the SM and chiral SM.

\item The values of the coupling constants found in the previous step
  are used as initial values for RG group running in the chiral phase
  of the EW theory. Since the parameters $\alpha_0$ and $\alpha_1$ are
  absent in the SM, the initial conditions for them are fixed to be
  zero.\footnote{The interpolation between the SM running of the
  coupling constants and the running in the chiral SM can be made
  smooth with the use of $q$-factor, defined in (\ref{qfactor}), in
  analogy with ref.~\cite{DeSimone:2008ei}. This leads, however, to
  negligible modifications of numerics, since $q(\chi)$ defined in Eq.
  (\ref{qfactor}) is a rapidly changing function in the region  $h
  \sim M_P/\xi$. A simple demonstration of this fact can be made by
  switching from the SM RG running to the chiral SM one at
  $M_P/\sqrt{\xi}$ instead of $M_P/\xi$.}

\item The two-loop effective potential is computed in the inflationary
  region as $U+U_1+U_2$, with all coupling constants taken at scale
  $\mu$, and the $Z$ boson mass given by (\ref{Zmass}).  To optimize
  the amplitude of higher order corrections, the parameter $\mu$ is
  chosen as in \cite{Bezrukov:2008ej}.

  For the renormalization prescription I the value of $\mu$ is
  \begin{equation}
    \label{renormcondition}
    \mu^2 = \kappa^2 m_t^2(\chi) =
    \kappa^2\frac{y_t(\mu)^2}{2}\frac{M_P^2}{\xi(\mu)} \left(
      1-e^{-\frac{2\chi}{\sqrt{6}M_P}} \right)
    \;.
  \end{equation}
  Here $\kappa$ is some constant of order one.\footnote{For technical
    reasons instead of expressions (\ref{renormcondition}),
    (\ref{renormconditionBarv}) we used for numerics the same
    formulas, but with the constants $y_t$, $\xi$ taken at the point
    $M_P/\xi$.}

  For the prescription II we take another value for $\mu$:
  \begin{equation}
    \label{renormconditionBarv}
    \mu^2 = m_t^2(\chi)\Omega(\chi)^2
    = \frac{y_t(\mu)^2}{2}\frac{M_P^2}{\xi(\mu)}
    \left(
      e^{\frac{2\chi}{\sqrt{6}M_P}}-1
    \right)
    \;.
  \end{equation}

\item As the potential is fixed, the value of $\xi$ is chosen in such
  a way that the correct WMAP5 normalization is reproduced. It is
  important to stress here that the wave-function normalization for
  the Higgs field in inflationary domain can be neglected, since the
  Higgs interactions with matter are suppressed by large value of
  non-minimal coupling $\xi$.

\item The spectral index $n_s$ and tensor-to scalar ratio $r$ are
  computed.  To be more specific, for the given potential the field
  $\chi$ corresponding to the end of inflation is found, then the
  field corresponding to 59 e-foldings of inflation is found, then at
  this scale $n_s$ and $r$ are calculated using standard slow-roll
  formulas.
\end{enumerate}

The downgrade of this procedure to tree mapping at $\mu=m_Z$, one-loop
running from this point to $M_P/\xi$, and the use of one-loop
effective potential in the inflationary region \emph{without} RG
improvement coincides with the computation carried out in our earlier
work \cite{Bezrukov:2008ej} up to higher-order terms in coupling
constants, due to Eq.~(18) of \cite{Bezrukov:2008ej}.\footnote{The RG
equations in the chiral phase of the EW theory found in the present
work are different from the RG equations used in
\cite{Bezrukov:2008ej} in this domain. However, the one-loop running
of the combination of couplings $\lambda/\xi^2$, most relevant for
determination of the inflationary potential, is exactly the same in
both phases, see Eq. (\ref{eq:lambdaxi}).}  The computation of
\cite{Barvinsky:2008ia} corresponds to the same procedure with the
prescription II but without RG running from $m_Z$ to $M_P/\xi$. In
\cite{Barvinsky:2008ia} the values of the coupling constants at $\mu =
M_P/\xi$ were taken to be the same as at $\mu=m_Z$. The computation
\cite{DeSimone:2008ei} corresponds nominally (for a more detailed
comparison see section~\ref{sec:comp}) to one-loop mapping at $\mu=m_t$
and the use of tree effective potential improved with the use of
two-loop RG equations up to the inflation scale.

\section{Numerical results}
\label{sec:num}

The results of the computation of the spectral index and
tensor-to-scalar ratio using the procedure from the previous section
with both prescriptions I and II are shown in Figs.~\ref{fig:ns},
\ref{fig:r}.  The running of the spectral index $dn_s/d\ln k$ is
always very small, of the order of $10^{-4}$.

\EPSFIGURE[t]{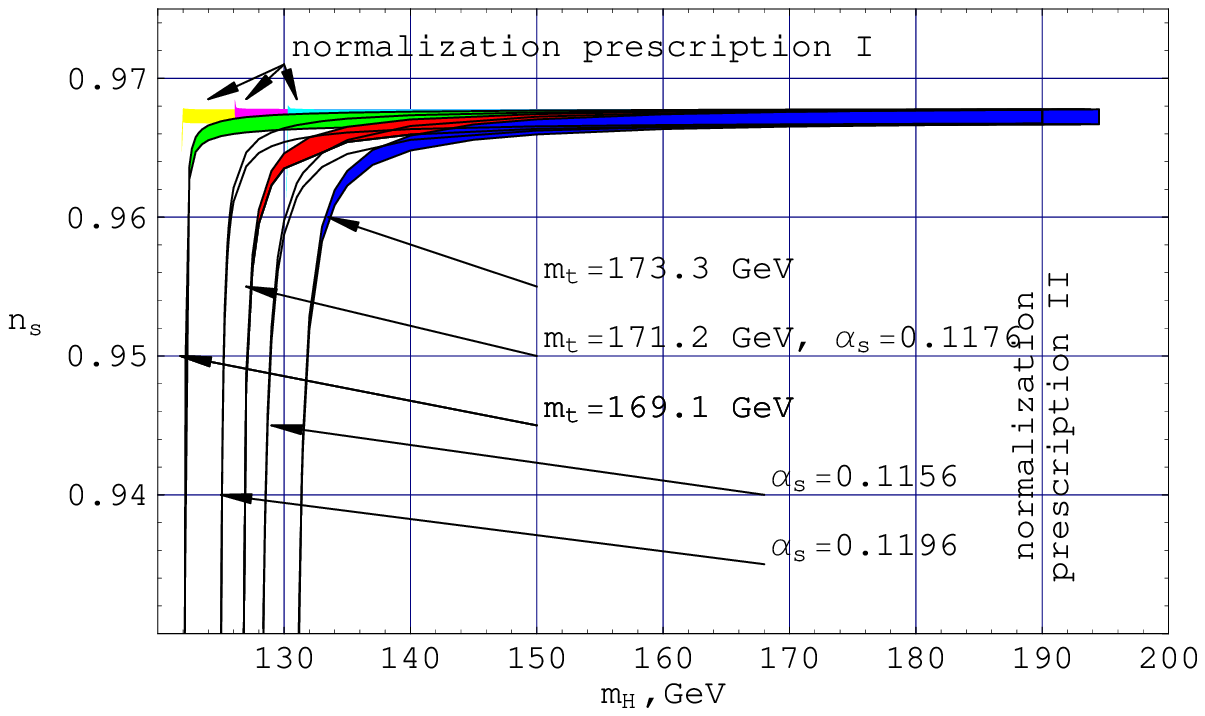}{Spectral index $n_s$ depending on the Higgs mass
  $m_H$, calculated with the RG enhanced effective potential.  Nearly
  horizontal coloured stripes correspond to the normalization
  prescription I and different $m_t$.  Green, red, and blue stripes
  give the result with normalization prescription II for different
  $m_t$ and $\alpha_s=0.1176$, two white regions correspond to
  different $\alpha_s$ and $m_t=\unit[171.2]{GeV}$.  The width of the
  stripes corresponds to changing the number of e-foldings between 58
  and 60, or approximately one order of magnitude in reheating
  temperature.
  \label{fig:ns}}

Let us first comment on the results in the prescription I.  The
spectral index $n_s$ is nearly constant, exactly the same as for
one-loop computation \cite{Bezrukov:2008ej}.  The only difference is the
shift of the minimal allowed Higgs boson masses to:
\begin{equation}
  \label{2loop}
  m^I_\mathrm{min} = \left[
    126.1 + \frac{m_t - 171.2}{2.1}\times4.1
    -\frac{\alpha_s-0.1176}{0.002}\times 1.5
  \right]\unit[\mbox{}]{GeV}
  \;,
\end{equation}
which happens mainly due to the two-loop running up to the $M_P/\xi$
scale.  One can see small wiggles at the lowest $m_H$ on the lines
(they were more pronounced in the one-loop result).  This corresponds
to the situation when $\lambda$ is approaching zero in the
inflationary region and the tree level part of the potential becomes
comparable with the loop corrections. It is unclear whether this
feature is a real physical effect or a result of approximation.

\FIGURE[t]{\centerline{%
    \includegraphics{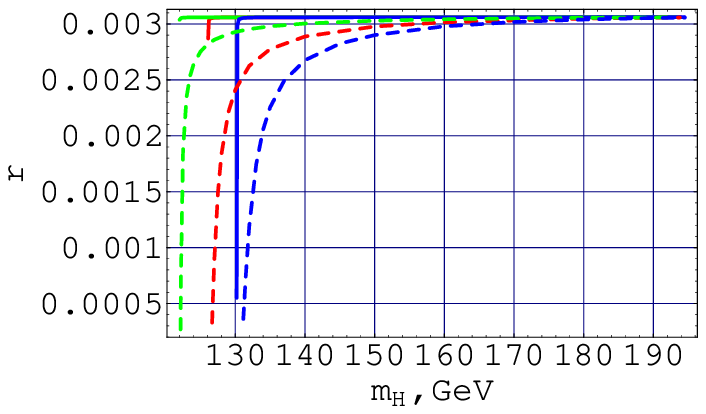}
    \caption{Tensor-to-scalar ratio $r$ depending on the Higgs
  mass $m_H$, calculated with the RG enhanced effective potential.
  Nearly horizontal solid lines correspond to the normalization
  prescription I.  Green, red, and blue dashed lines give the result
  with normalization prescription II for
  $m_t=169.1,171.2,\unit[173.3]{GeV}$.  Dependence on the number of
  e-foldings is very small.
  \label{fig:r}
}}}
At the high $m_H$, contrary to the one-loop analysis, the solution to
two-loop RG equations for the scalar self-coupling does not show up a
Landau-pole behaviour.  Though $\lambda$ enters into strong-coupling
region, it stays finite.  Of course, perturbation theory cannot be
trusted in this case.  We believe that inflation is hardly possible in
this situation.  The lattice simulations of pure scalar theory
\cite{Luscher:1987ay} show that the corresponding theory has nothing
in common with a continuum strongly coupled field theory. Requiring
that one should have the weak coupling (somewhat arbitrary we set the
requirement $\lambda <6$) in the region $M_P/\xi$ we obtain, in 2-loop
approximation,
\begin{equation}
  \label{2loop-max}
  m^I_\mathrm{max} = \left[
    193.9
    + \frac{m_t - 171.2}{2.1}\times0.6
    -\frac{\alpha_s-0.1176}{0.002}\times 0.1
  \right]\unit[\mbox{}]{GeV}
  \;.
\end{equation}

Note that the requirement to have weak coupling up to $\mu = M_P$ is
somewhat stronger, $m_H < \unit[173.5]{GeV}$ (for central values of
$m_t$ and $\alpha_s$). Also, the requirement that $\lambda(\mu) >0$ for
$\mu <M_P$ is just slightly stronger, than (\ref{2loop}), giving the
interval of the Higgs masses for which the SM remains a good quantum
field theory all the way up to the Planck scale, see
Fig.~\ref{fig:lambdarun}.

Figures \ref{fig:xi} and \ref{fig:lambda} give for reference the
values of $\xi$ and $\lambda$ at the scale $M_P/\xi$.  One can see,
that the tree level COBE normalization requirement
\cite{Bezrukov:2008ut} holds very well for high Higgs masses, and
becomes violated at lower ones, see Fig.~\ref{fig:cobe}.

For prescription II the behaviour is qualitatively similar to the
one-loop calculation in \cite{Bezrukov:2008ej}.  Numerically, we have
the overall shift in the Higgs mass region of $\unit[10]{GeV}$,
corresponding to the difference between one- and two-loop running of
the coupling constants up to the inflationary region $M_P/\xi$.  Also,
the deviation from the asymptotic value of the spectral index $n_s$ at
small Higgs masses is smaller, than given by the one loop calculation.
For low Higgs masses the spectral index drops, and goes
out of the allowed region ($n_s>0.93$) for
\begin{equation}
  \label{2loop-II}
  m^{II}_\mathrm{min} = \left[
    126.7
    + \frac{m_t - 171.2}{2.1}\times4.5
    - \frac{\alpha_s-0.1176}{0.002}\times 1.7
  \right] \unit[\mbox{}]{GeV}
  \;.
\end{equation}
This is a bit stronger, than (\ref{2loop}).

\EPSFIGURE[t]{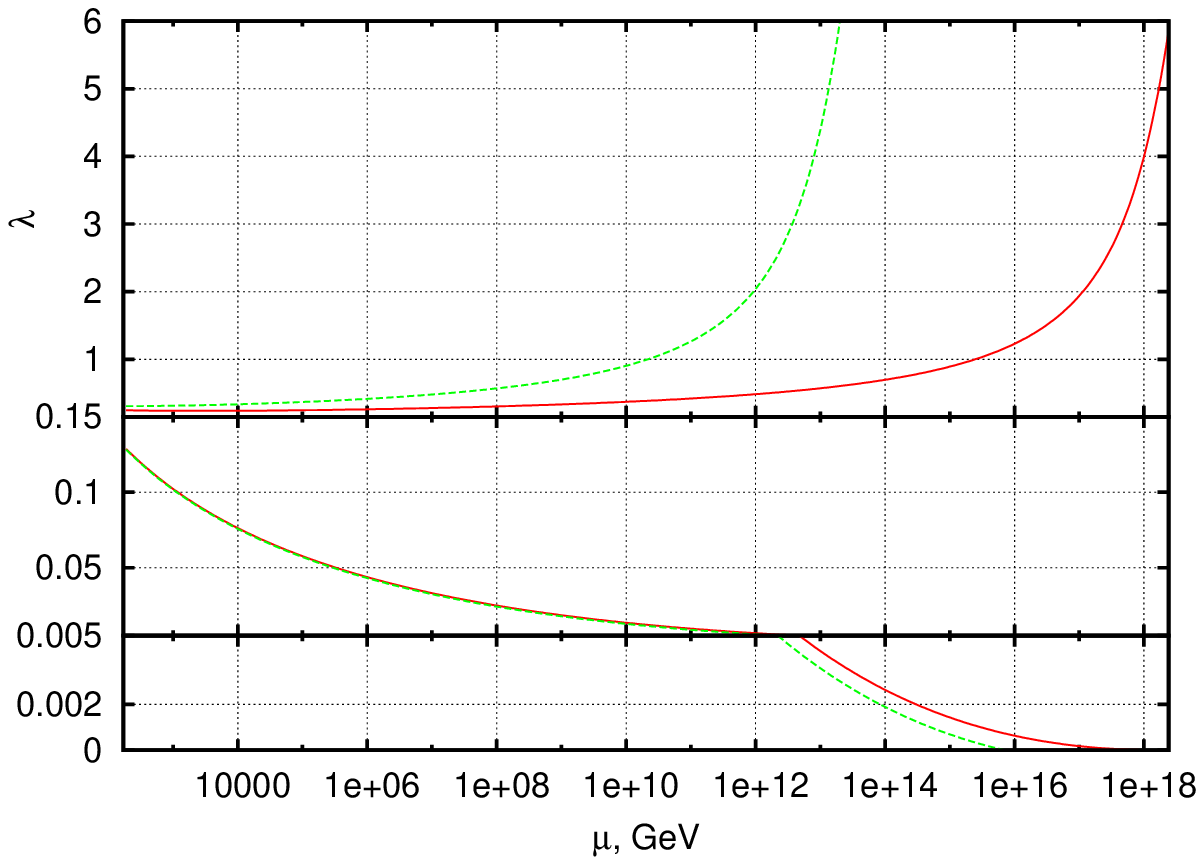}{The running of the Higgs coupling constant
  $\lambda$ up to the high scale in the canonical SM.  Solid line
  graphs correspond to $m_H=\unit[174]{GeV}$ and
  $m_H=\unit[126.3]{GeV}$, leading to $\lambda=6$ at $M_P$, or
  $\lambda$ touching $0$ at a scale slightly below (these values
  represent the window in which the SM is valid all the way to the
  Planck scale, for central values of $m_t$ and $\alpha_s$).  Dashed
  graphs correspond to $m_H=\unit[194]{GeV}$ and
  $m_H=\unit[126.1]{GeV}$, leading to strong coupling or zero at
  $M_P/\xi$ (for $\xi=11600$ and $704$, respectively).  The lower part
  of the graph is scaled in the vertical direction for visibility.
  \label{fig:lambdarun}
}

\DOUBLEFIGURE{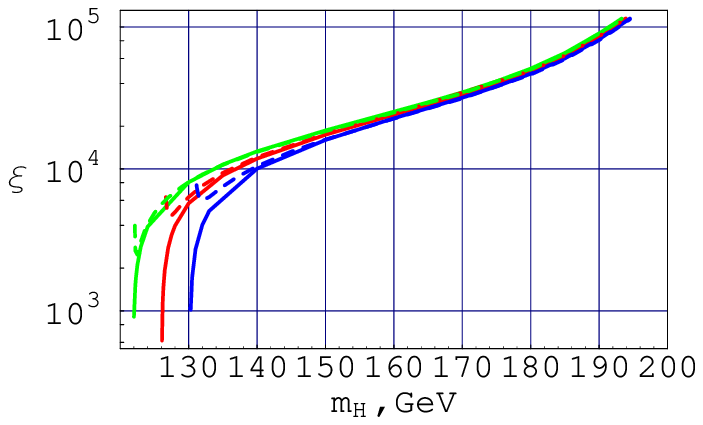}{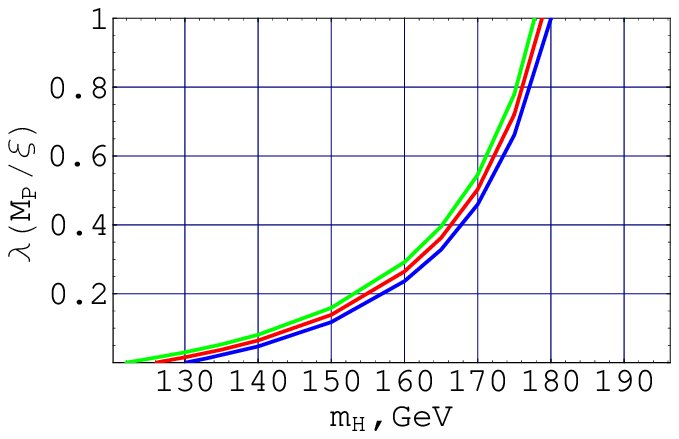}{$\xi$ at the scale $M_P/\xi$ depending
  on the Higgs mass $m_H$ for $m_t=\unit[169.1,171.2,173.3]{GeV}$
  (from upper to lower graph).  Solid lines correspond to prescription
  I, dashed~--- to prescription II.  Changing the e-foldings number and
  error in the WMAP normalization measurement introduce changes
  invisible on the graph.
  \label{fig:xi}}{$\lambda$ at the scale $M_P/\xi$ depending on the
  Higgs mass $m_H$ for $m_t=\unit[169.1,171.2,173.3]{GeV}$ (from upper
  to lower graph).  Changing the e-foldings number and
  error in the WMAP normalization measurement introduce changes
  invisible on the graph.
  \label{fig:lambda}}
\EPSFIGURE{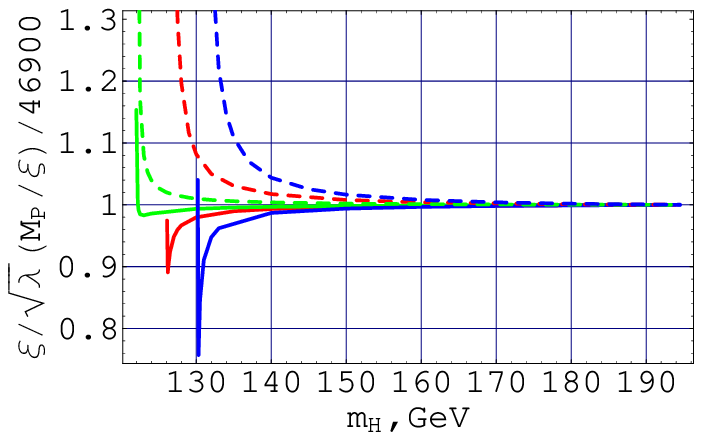}{The ratio $\xi/\sqrt{\lambda}/46900$ at the
  scale $M_P/\xi$.  The ratio equal to one \cite{Bezrukov:2008ut}
  corresponds to the COBE normalization requirement for tree level
  potential.  Solid lines are for prescription I,
  dashed~--- prescription II.  Green, red and blue lines (from left to
  right) correspond to $m_t=\unit[169.1,171.2,173.3]{GeV}$.
  \label{fig:cobe}}

It is interesting to note, that at large Higgs masses in any
calculation the spectral index approaches the tree level value
$n_s=0.967$.  The reason is that radiative corrections play a
subdominant role if $\lambda$ is not very small, and the running of
the coefficient $\lambda/\xi^2$ in front of the tree level potential
is less important for large $\lambda$ (see Eq.~(\ref{eq:lambdaxi})).

The comparison between different approximations for the effective
potential can be found in Fig.~\ref{fig:ns-all}.  For this figure we
always made two-loop SM running of the constants up to the $M_P/\xi$
scale, to evade the overall mass shift and allow for comparison.
Then, we plotted one-loop effective potential (\ref{U1loop}) and
two-loop potential (\ref{egeneral}) \emph{without} any further RG
improvement; tree level potential with constants improved by chiral RG
running up to the scale (\ref{renormcondition}) (or
(\ref{renormconditionBarv}) for prescription II), and RG enhanced tree
level potential plus one- or two-loop potential.

In prescription I all results are almost indistinguishable, because
the normalization scale (\ref{renormcondition}) changes very little
during the inflation.

In prescription II the results differ.  We can see, that addition of
the two-loop corrections changes the one-loop result rather strongly,
because of accidental cancellation of the one-loop contributions from
the gauge bosons and the top quark at one-loop.  The RG enhanced
result is rather close to the result obtained by using the two-loop
effective potential.  Furthermore, adding loop corrections to the RG
enhanced tree-level result does not change the results much.

Our results are stable against variation of the parameter $\kappa$,
defined in (\ref{renormcondition}), (\ref{renormconditionBarv}),
demonstrating that the running of the coupling constants is indeed
compensated by the logs in the effective potential (as it should be).
In Fig.~\ref{fig:Ukappa} we show the change of the different
approximations to effective potential due to the change of $\kappa$.
It is seen, that adding one- and two-loop logarithms to the
RG-enhanced tree level potential removes the superficial $\mu$
dependence of the potential.

The close similarity of the one-loop and two-loop results demonstrates
that our conclusions are stable against radiative corrections for both
renormalization procedures. The main effect of accounting of two-loop
running and one-loop matching is the shift of the admissible interval
for Higgs boson masses. However, the spectral index $n_s$ cannot be
determined unambiguously in the SM inflation by giving the exact
values of the SM parameters (in particular the Higgs and top
masses). Though $n_s$ remains close to its tree value in the wide
range of Higgs masses, (small) deviations from it contain an extra
uncertainty, coming from the choice of the normalization point (cf.\
prescriptions I and II discussed in section~\ref{sec:procedure}).
From theoretical point of view, this uncertainty can only be removed
if the physics at Planck scales is fixed. A more exact cosmological
measurement of $n_s$, together with precise knowledge of the SM
parameters can help to resolve this issue.

It is interesting to note that the requirement of sucessful
Higgs-inflation allows to put an upper bound on the mass of the top
quark.  If t-quark mass were larger than \unit[240-250]{GeV}, no
choice of $\xi$ and of the Higgs mass could lead to acceptable
inflationary parameters. This number is smaller than the experimental
lower constraint on the mass of the $t'$-quark of the fourth
generation \cite{Amsler:2008zzb}, meaning that Higgs-inflation in the
SM is not consistent with fourth fermionic family.

We conclude this section with an estimate of the theoretical
uncertainties related to higher order loop corrections\footnote{We
thank Mikhail Kalmykov for discussion of this issue.}. To test the
accuracy of the pole matching procedure at low energy scale one can
use somewhat different procedures to compute the values of the
coupling constants at $\mu=m_t$. First, we simply used the formulas
given in \cite{Kajantie:1995dw,Espinosa:2007qp} for $\mu=m_t$.  Then
we used the same equations from \cite{Kajantie:1995dw} at $\mu=m_Z$,
got the coupling constants at this point, and then considered their
running up to $\mu=m_t$. The difference in the values gives an
estimate of the uncertainties related to the two-loop terms in the
matching procedure.  For small Higgs masses in the region $m_H \simeq
\unit[130]{GeV}$ the variation in the scalar self-coupling
is\footnote{We give two significant digits for uncertainties in
$\lambda$ and $y_t$ in order to allow for a detailed comparison.}
$\delta\lambda(m_t)/\lambda(m_t) \simeq 0.018$ and in the top Yukawa
coupling $\delta y_t(m_t)/y_t(m_t) \simeq 0.0015$ (uncertainties in
other couplings are much smaller). This leads to the effective
redefinition of the Higgs mass by $\unit[1.2]{GeV}$ and the top mass
by $\unit[0.3]{GeV}$. The change of the top mass by this amount leads
to uncertainty in the lower limit on the Higgs mass equal to
$\unit[0.6]{GeV}$. The three-loop $\alpha_s$ correction to the top
mass  gives $\delta y_t(m_t)/y_t(m_t) \simeq 0.0042$
\cite{Chetyrkin:1999qi,Melnikov:2000qh}, the non-perturbative QCD
effects in the top pole mass - \MSb{} mass matching are expected to be
at the level of  $\delta y_t(m_t)/y_t(m_t) \simeq 0.001$
\cite{Smith:1996xz}. The four-loop $\alpha_s$ contribution to the top
mass was guessed to be the same \cite{Kataev:2009ns}. If we assume
that these uncertainties are not correlated, and symmetric, we get a
theoretical error in the determination of the critical Higgs mass,
$\delta m_\mathrm{theor} \simeq \unit[2.2]{GeV}$. This error is
smaller than the present uncertainties related to the experimental
errors in the top mass and in the strong coupling constant. It can be
reduced further to $\delta m_\mathrm{theor} \simeq \unit[0.4]{GeV}$ by
the complete two-loop pole matching (not known at the moment),
accounting for 3-loop $\alpha_s$ correction to $y_t$, found in
\cite{Chetyrkin:1999qi,Melnikov:2000qh}, and by the three-loop RG
running up to the scale $M_P/\xi$ (note that the complete 3-loop
$\beta$-functions for the Standard Model are not known yet).

\DOUBLEFIGURE{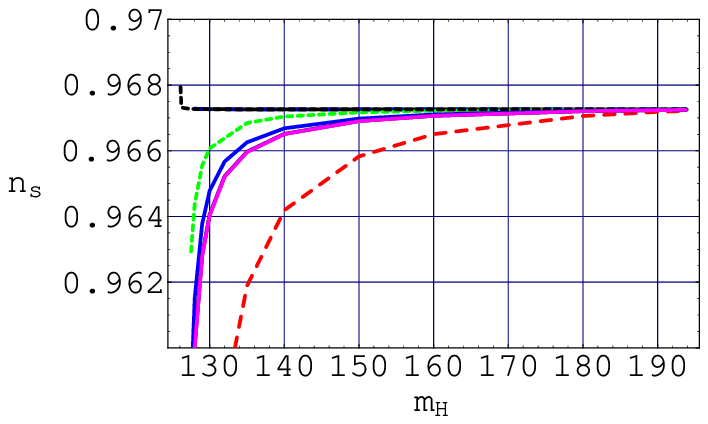}{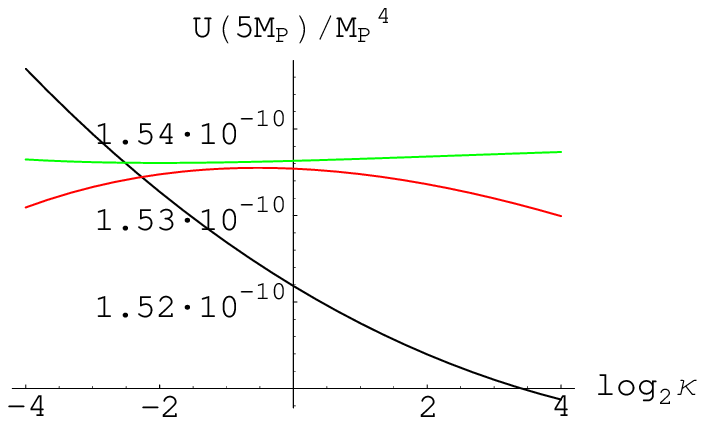}{Spectral index $n_s$ depending on the
  Higgs mass $m_H$ calculated in different approximations.  Nearly
  horizontal line represents the result in the prescription I
  (different approximations lead to indistinguishable results).
  Everything else is the results in prescription II.  Red dashed line
  (lower) is the one loop effective potential, green dashed line
  (upper) is the two loop effective potential, close solid lines are
  RG enhanced results without (blue) and with addition of the one or
  two loop effective potential (red and black, nearly
  indistinguishable).  For all graphs $m_t=\unit[171.2]{GeV}$,
  $\alpha_s=0.1176$.
  \label{fig:ns-all}}{Result of moving the normalization point
  (\protect\ref{renormcondition}) in the potential.  The black falling
  graph is the tree level contribution, red (lower) and green (upper)
  demonstrate removal of the dependence on the normalization point.
  Potential is calculated in prescription I for the typical
  inflationary field $\chi=5M_P$ and $m_H=\unit[140]{GeV}$.
  \label{fig:Ukappa}}

\section{Comparison with refs.~\cite{Bezrukov:2008ej,DeSimone:2008ei}}
\label{sec:comp}
Our preprint \cite{Bezrukov:2008ej} appeared in hep-ph simultaneously
with the paper \cite{DeSimone:2008ei}, devoted to the study of the
same problem.  The analysis of \cite{DeSimone:2008ei} was carried out
with the use of two-loop RG equations for the running couplings,
modified to account for non-minimal coupling of the Higgs field to
Ricci scalar in the Jordan frame. In \cite{Bezrukov:2008ej} we used
one-loop approximation and performed the computations of the effective
potential in the Einstein frame.

In very general terms, the findings of \cite{Bezrukov:2008ej}, the
present paper and of \cite{DeSimone:2008ei} are quite similar.
Namely, in \cite{Bezrukov:2008ej} it was shown that the
Higgs-inflation in the SM, consistent with WMAP observations, is
possible provided the Higgs mass $m_H$ lies in the interval
$m_\mathrm{min}< m_H < m_\mathrm{max}$, somewhat exceeding the region of
the Higgs masses in which the SM is a viable effective field theory up
to the Planck scale. In one-loop approximation the minimal and maximal
Higgs masses were found to be
\begin{align}
  \label{mhiggs}
  m_\mathrm{min} =& [136.7 + (m_t - 171.2)\times1.95]\unit[\mbox{}]{GeV}\;,\\
  m_\mathrm{max} =& [184.5 + (m_t - 171.2)\times0.5]\unit[\mbox{}]{GeV}\;.
  \nonumber
\end{align}
The uncertainty resulting from variation of the strong coupling
constant was not estimated in \cite{Bezrukov:2008ej}.

With the use of two-loop analysis, the paper \cite{DeSimone:2008ei}
found that the successful SM inflation can take place if the Higgs
mass exceeds the value
\begin{align}
  m_\mathrm{min} = [125.7 + \frac{m_t - 171}{2}\times3.8
  -\frac{\alpha_s-0.1176}{0.0020}\times 1.4]~\unit{GeV}\pm \delta
  \;,
\end{align}
where $\delta \sim \unit[2]{GeV}$ indicates theoretical uncertainty from
higher order corrections.

At the quantitative level, the one-loop value of
\cite{Bezrukov:2008ej} for $m_\mathrm{min}$ is some $\unit[11]{GeV}$
larger that that of \cite{DeSimone:2008ei}. The result of our present
paper for the minimal allowed Higgs mass is given in
eqns.~(\ref{2loop}), (\ref{2loop-II}) and almost coincides with the
bound found in \cite{DeSimone:2008ei}.  Therefore, the shift of the
critical Higgs mass is associated with the account for one-loop
corrections at the low-energy matching (instead of tree procedure) and
with two-loop (instead of the one-loop) running from the low scale to
the high scale. The existence of this shift due to higher order
effects was anticipated in \cite{Bezrukov:2008ej}.

The behaviour of the spectral index as a function of the Higgs mass is
a more subtle effect.  The choice of the normalization point of
\cite{DeSimone:2008ei} coincides with that of \cite{Barvinsky:2008ia}
and corresponds to the prescription II of \cite{Bezrukov:2008ej} and
of the present work. Let us compare our results  for this case with
those of \cite{DeSimone:2008ei}.

According to \cite{DeSimone:2008ei}, if $m_H$ decreases, $n_s$
increases and is always larger than the tree value (see Fig.~1 of
\cite{DeSimone:2008ei}). According to computations of
\cite{Bezrukov:2008ej} and of the present work, $n_s$ decreases and is
always smaller than the tree value (Fig.~3 of \cite{Bezrukov:2008ej}
and our Fig.~\ref{fig:ns}).  The similar type of discrepancy
exists for the behaviour of the tensor-to-scalar ratio (cf.\ Fig.~5 of
\cite{DeSimone:2008ei} and our Fig.~\ref{fig:r}).  The two-loop computation
carried out in the present work shows that the difference between the
results of \cite{DeSimone:2008ei} and \cite{Bezrukov:2008ej} is
\emph{not} related to the number of loops accounted for. Below we will
try to elucidate possible origins of this discrepancy. Making a
detailed comparison of computations in both articles is difficult as
we used different frames for analysis and solved different equations.
However, there are several essential points which are treated
differently in these works even in one-loop approximation.

\begin{description}
\item[(I)] The first point is related to gauge invariance. Our method (and
the one used in \cite{Bezrukov:2008ej}) is explicitly gauge-invariant.
The $\beta$-functions for RG equations we used do not depend on the
gauge choice, and the running of the couplings up to the point
$M_P/\xi$ is gauge independent. The boundary conditions are
gauge-invariant as well. The method of \cite{DeSimone:2008ei} includes
explicitly the gauge non-invariant object~--- the anomalous dimension of
the scalar field $\gamma$. The equation (A.5) of this work indicates
that the Landau gauge was used (in the arbitrary $\alpha$-gauge one
would get
\be
\gamma=\frac{1}{16\pi^2}\left((2+\alpha)\frac{3g^2+g'^2}{4}
-3y_t^2\right)+\dots
\ee
(here $\alpha$ is the coefficient in front of the longitudinal term of
the gauge-field propagator,  $\alpha=1$ and $\alpha=0$ correspond to
the Landau and Feynman gauges respectively). The gauge non-invariant
parameter $\gamma$ is used then for RG running of all coupling
constant through equations like
$d\lambda/dt=\beta_\lambda/(1+\gamma)$. Clearly, the definition of all
couplings is then not gauge-invariant. It does not correspond to the
standard gauge-invariant \MSb{} prescription. Therefore, the pole-\MSb{}
matching described in \cite{Espinosa:2007qp}  is not
appropriate for fixing the initial condition for the \emph{gauge
non-invariant} couplings used in \cite{DeSimone:2008ei}. The gauge
variation of the scalar self-coupling $\lambda$ (most important for
computation of inflationary spectral indexes) reads
\be
\delta\lambda = -\frac{\delta\alpha}{16\pi^2}\int^t
dt\beta(\lambda) \left(\frac{3g^2+g'^2}{4}\right)\;,
\ee
where $\delta\alpha$ is the change of the gauge parameter. It does
introduce the gauge dependence of the results at two-loop level and
may result in the change of $n_s$ behaviour with the Higgs mass.

\item[(II)] The second point is related to consistency of introducing of
$s$-factor, which accounts for gravity-Higgs mixing in the Jordan
frame. One can see that the argument leading to Eq.~(4.6) of
\cite{DeSimone:2008ei} is only applicable to the physical Higgs field
and does not work for Nambu-Goldstone bosons present in the Landau
gauge, used in \cite{DeSimone:2008ei}. Therefore, the kinetic term for
these fields is not modified. In other words, the virtual
Nambu-Goldstone bosons, contributing to diagrams for $\beta$-functions
must be kept.  This fact was accounted for in our paper
(section~\ref{sec:rg}) but not in \cite{DeSimone:2008ei}, as the
comparison of the one-loop parts of relations (A.1) and (A.2) and our
equations in section~\ref{sec:rg} shows. One finds the differences in
the running of the SU(2)$\times$U(1) gauge couplings $g$ and $g'$, of
the scalar self-coupling $\lambda$ and of the top Yukawa coupling
$y_t$. For example, the $\beta$-function for $y_t$ does not contain
any contribution from $y_t$ in the inflationary region in
\cite{DeSimone:2008ei}, whereas it is in fact present. We suspect that
this may be one of the reasons for the opposite behaviour of the
spectral index $n_s$ found in \cite{DeSimone:2008ei}, which
underestimates this contribution. Moreover, the running of extra
couplings $\alpha_0$ and $\alpha_1$, necessary for description of the
electroweak theory in the inflationary region, was not considered in
\cite{DeSimone:2008ei}.

\item[(III)] Yet another (numerically less important) difference is related
to accuracy of computation. Reference~\cite{DeSimone:2008ei} did not
compute one- or two-loop effective potential but used the tree
potential improved by the renormalization group. In this way the
leading logs are accounted for, but the terms without logs are lost.
In our present work all these terms up to the two-loop order are
included.
\end{description}

The second version of \cite{DeSimone:2008ei} contains a ``Note Added''
in which a number of remarks about our work \cite{Bezrukov:2008ej} has
been made. We comment on them below. For the reader convenience, we
quote here the most relevant part of the ``Note Added'' of
ref.~\cite{DeSimone:2008ei}:

``In our analysis, we computed the full RG improved effective
potential. We did this including (i) 2-loop beta functions, (ii) the
effect of curvature in the RG equations (through the function $s$),
(iii) wavefunction renormalization, and (iv) accurate specification of
the initial conditions through proper pole matching. On the other
hand,  \cite{Bezrukov:2008ej} did not compute the full effective
potential or include any of the items (i)--(iv).\footnote{Though
wavefunction renormalization was not included in
\cite{Bezrukov:2008ej}, external leg corrections in the running of
$\lambda$ were included. However these two effects roughly cancel
against one another.} Instead ref.~\cite{Bezrukov:2008ej}
approximated the potential at leading log order with couplings
evaluated at an inflationary scale after running them at 1-loop (this
is one step beyond  \cite{Barvinsky:2008ia} where couplings were not
run).''

We certainly agree with \cite{DeSimone:2008ei} that in
\cite{Bezrukov:2008ej} we did not include (i) 2-loop $\beta$ functions
(we had one-loop running) and (iv) one-loop pole matching for initial
conditions for renormalization group (we used the tree values). We
said explicitly that our aim is the \emph{one-loop analysis} of the
problem, for which the tree pole matching is adequate. As for the
point (ii), the \emph{one-loop} renormalization group running of the
combination $\lambda/\xi^2$, relevant for inflation, is \emph{exactly
the same} at small and at large Higgs field values, meaning that our
analysis is perfectly consistent at one loop. Concerning the point
(iii), our formalism \emph{does not} require the computation of the
wave-function renormalization. At the same time, the running of $\xi$
(which was not accounted for in \cite{DeSimone:2008ei}) plays an
important role. See also the discussion of wave-function
renormalization and pole-matching procedure in (I), and a comment on
``full RG improved effective potential'' in (III).

\section{Conclusions}
\label{sec:conc}

Inflation does not necessarily require the existence of the
inflaton~--- the Higgs boson of the Standard Model can make the Universe
flat, homogeneous and isotropic, and produce fluctuations necessary
for structure formation. This happens in the SM with sufficiently
large non-minimal coupling of the Higgs to the gravitational Ricci
scalar.  An important requirement is the validity of the SM all the
way up to the Planck scale. In this case radiative corrections to the
inflationary potential can be computed. They can be used to put
constraints on the Higgs mass and to determine the connection between
the cosmological parameters and properties of particles in the
Standard Model. The window of allowed masses for the Higgs boson is
$m_H\in[126,194]\unit[\mbox{}]{GeV}$ (for central values of $m_t$ and
$\alpha_s$ and neglecting theoretical uncertaites). It somewhat
exceeds the region where the canonical SM is a valid field theory up
to the Planck scale, $m_H\in[126.3,174]\unit[\mbox{}]{GeV}$.  This
prediction can be testable at the LHC. The precision measurements of
$n_s$ and $r$, together with exact knowledge of the SM parameters will
allow to shed light on the origin of inflation and on Planck scale
physics.

\acknowledgments

The authors thank Sergei Sibiryakov, Andrei Barvinsky and Dmitry
Gorbunov for valuable discussions.  The exchange of information and
discussions with Andrea De Simone, Mark Hertzberg and Frank Wilczek
are appreciated.  This work was supported by the Swiss National
Science Foundation.

\appendix

\section{Initial conditions}

\FIGURE{\centerline{%
    \includegraphics{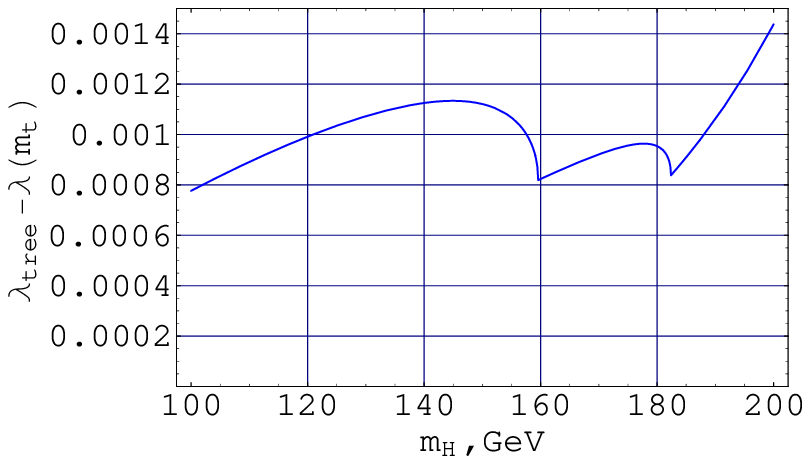}
    \caption{Correction to the Higgs self coupling
      $\delta_h=\lambda_\mathrm{tree}-\lambda(m_t)$ depending on the Higgs
      mass $m_H$.
      \label{fig:lambdadiff}}}}

The \MSb{} values for the gauge coupling constants at $\mu=m_t$ were
obtained from the $\mu=m_Z$ present in \cite{Amsler:2008zzb} by
one-loop running for $g'$, $g$ and two-loop running for $g_s$.  The
resulting values for the gauge couplings at $m_t$ are $g'=0.3587$,
$g=0.6484$, $g_s=1.1619 + 4.5(\alpha_s - 0.1176)$.  Close values for
electroweak constants were also obtained directly from the pole masses
using formulas from \cite{Kajantie:1995dw}.

The top quark Yukawa constant $y_t$ and Higgs self-interaction
$\lambda$ at $\mu=m_t$ can be obtained using the expressions from
appendix of \cite{Espinosa:2007qp}.  To summarize,
the \MSb{} value for the top Yukawa is very well approximated as
\begin{align}
  y_t(m_t) =& \frac{\sqrt{2}m_t}{v}(1+\delta_t)\;,\\
  \delta_t=&-0.0566
  + \left(\frac{m_t - \unit[171.2]{GeV}}{\unit[6500]{GeV}}\right)
  + \left(\frac{\alpha_S(m_Z)-0.1176}{2}\right)
  - \left(\frac{m_H - \unit[130]{GeV}}{\unit[55000]{GeV}}\right)
  \;,\nonumber
\end{align}
where $v^2\equiv\frac{1}{\sqrt{2}G_F}=(\unit[246.221]{GeV})^2$.  The
initial data for $\lambda$ is $\lambda(m_t)=m_H^2/2v^2-\delta_h$,
where $\delta_h$ is given in Fig.~\ref{fig:lambdadiff}.

Using formulas from \cite{Kajantie:1995dw} similar values are
obtained. The main difference in $\delta_t$ comes from the two-loop
QCD contributions accounted for in \cite{Espinosa:2007qp} and not in
\cite{Kajantie:1995dw}. The difference in $\delta_h$ comes from
neglecting the W and Z mass difference in the loop integrals in
\cite{Kajantie:1995dw}, but accounted in \cite{Espinosa:2007qp}.
Numerically, this just slightly changes the influence of the top mass
on the overall results.

\bibliographystyle{JCAP-hyper}
\bibliography{refs}

\providecommand{\href}[2]{#2}\begingroup\raggedright\begin{thebibliography}{10}

\bibitem{Bezrukov:2008ej}
F.~L. Bezrukov, A.~Magnin, and M.~Shaposhnikov, {\it {Standard Model Higgs
  boson mass from inflation}},
  \href{http://dx.doi.org/10.1016/j.physletb.2009.03.035}{{\em Phys. Lett.}
  {\bf B675} (2009)  88--92}, [\href{http://xxx.lanl.gov/abs/0812.4950}{{\tt
  arXiv:0812.4950}}],
  \href{http://www.slac.stanford.edu/spires/find/hep/www?rawcmd=eprint+0812.49%
50}{[SPIRES]}.

\bibitem{DeSimone:2008ei}
A.~De~Simone, M.~P. Hertzberg, and F.~Wilczek, {\it {Running Inflation in the
  Standard Model}},  \href{http://xxx.lanl.gov/abs/0812.4946}{{\tt
  arXiv:0812.4946}},
  \href{http://www.slac.stanford.edu/spires/find/hep/www?rawcmd=eprint+0812.49%
46}{[SPIRES]}.

\bibitem{Linde:2007fr}
A.~Linde, {\it Inflationary cosmology},
  \href{http://dx.doi.org/10.1007/978-3-540-74353-8_1}{{\em Lect. Notes Phys.}
  {\bf 738} (2008)  1--54}, [\href{http://xxx.lanl.gov/abs/0705.0164}{{\tt
  arXiv:0705.0164}}],
  \href{http://www.slac.stanford.edu/spires/find/hep/www?rawcmd=eprint+0705.01%
64}{[SPIRES]}.

\bibitem{Mukhanov:2005sc}
V.~Mukhanov, {\em Physical foundations of cosmology}.
\newblock Cambridge, UK: Univ. Pr., 2005.
\newblock 421 p.

\bibitem{Komatsu:2008hk}
{\bf WMAP} Collaboration, E.~Komatsu {\em et.~al.}, {\it {Five-Year Wilkinson
  Microwave Anisotropy Probe (WMAP) Observations:Cosmological Interpretation}},
   \href{http://dx.doi.org/10.1088/0067-0049/180/2/330}{{\em Astrophys. J.
  Suppl.} {\bf 180} (2009)  330--376},
  [\href{http://xxx.lanl.gov/abs/0803.0547}{{\tt arXiv:0803.0547}}],
  \href{http://www.slac.stanford.edu/spires/find/hep/www?rawcmd=eprint+0803.05%
47}{[SPIRES]}.

\bibitem{Bezrukov:2007ep}
F.~L. Bezrukov and M.~Shaposhnikov, {\it {The Standard Model Higgs boson as the
  inflaton}},  \href{http://dx.doi.org/10.1016/j.physletb.2007.11.072}{{\em
  Phys. Lett.} {\bf B659} (2008)  703--706},
  [\href{http://xxx.lanl.gov/abs/0710.3755}{{\tt arXiv:0710.3755}}],
  \href{http://www.slac.stanford.edu/spires/find/hep/www?rawcmd=eprint+0710.37%
55}{[SPIRES]}.

\bibitem{Spokoiny:1984bd}
B.~L. Spokoiny, {\it Inflation and generation of perturbations in broken
  symmetric theory of gravity},
  \href{http://dx.doi.org/10.1016/0370-2693(84)90587-2}{{\em Phys. Lett.} {\bf
  B147} (1984)  39--43},
  \href{http://www.slac.stanford.edu/spires/find/hep/www?rawcmd=j+PHLTA,B147,3%
9}{[SPIRES]}.

\bibitem{Salopek:1988qh}
D.~S. Salopek, J.~R. Bond, and J.~M. Bardeen, {\it Designing density
  fluctuation spectra in inflation},
  \href{http://dx.doi.org/10.1103/PhysRevD.40.1753}{{\em Phys. Rev.} {\bf D40}
  (1989)  1753},
  \href{http://www.slac.stanford.edu/spires/find/hep/www?rawcmd=j+PHRVA,D40,17%
53}{[SPIRES]}.

\bibitem{Fakir1990}
R.~Fakir and W.~G. Unruh, {\it Improvement on cosmological chaotic inflation
  through nonminimal coupling},
  \href{http://dx.doi.org/10.1103/PhysRevD.41.1783}{{\em Phys. Rev.} {\bf D41}
  (1990)  1783--1791},
  \href{http://www.slac.stanford.edu/spires/find/hep/www?rawcmd=j+PHRVA,D41,17%
83}{[SPIRES]}.

\bibitem{Komatsu:1999mt}
E.~Komatsu and T.~Futamase, {\it Complete constraints on a nonminimally coupled
  chaotic inflationary scenario from the cosmic microwave background},  {\em
  Phys. Rev.} {\bf D59} (1999)  064029,
  [\href{http://xxx.lanl.gov/abs/astro-ph/9901127}{{\tt astro-ph/9901127}}],
  \href{http://www.slac.stanford.edu/spires/find/hep/www?rawcmd=eprint+astro-p%
h\%2F9901127}{[SPRIES]}.

\bibitem{Tsujikawa:2004my}
S.~Tsujikawa and B.~Gumjudpai, {\it Density perturbations in generalized
  einstein scenarios and constraints on nonminimal couplings from the cosmic
  microwave background},  {\em Phys. Rev.} {\bf D69} (2004)  123523,
  [\href{http://xxx.lanl.gov/abs/astro-ph/0402185}{{\tt astro-ph/0402185}}],
  \href{http://www.slac.stanford.edu/spires/find/hep/www?rawcmd=eprint+astro-p%
h\%2F0402185}{[SPRIES]}.

\bibitem{Barvinsky1994}
A.~O. Barvinsky and A.~Y. Kamenshchik, {\it Quantum scale of inflation and
  particle physics of the early universe},
  \href{http://dx.doi.org/10.1016/0370-2693(94)91253-X}{{\em Phys.Lett. B} {\bf
  332} (1994)  270--276}, [\href{http://xxx.lanl.gov/abs/gr-qc/9404062}{{\tt
  gr-qc/9404062}}],
  \href{http://www.slac.stanford.edu/spires/find/hep/www?rawcmd=eprint+gr-qc\%%
2F9404062}{[SPRIES]}.

\bibitem{Barvinsky1998}
A.~O. Barvinsky and A.~Y. Kamenshchik, {\it Effective equations of motion and
  initial conditions for inflation in quantum cosmology},
  \href{http://dx.doi.org/10.1016/S0550-3213(98)00484-2}{{\em Nucl.Phys. B}
  {\bf 532} (1998)  339--360},
  [\href{http://xxx.lanl.gov/abs/hep-th/9803052}{{\tt hep-th/9803052}}],
  \href{http://www.slac.stanford.edu/spires/find/hep/www?rawcmd=eprint+hep-th%
\%2F9803052}{[SPRIES]}.

\bibitem{Bezrukov:2008ut}
F.~Bezrukov, D.~Gorbunov, and M.~Shaposhnikov, {\it {On initial conditions for
  the Hot Big Bang}},  \href{http://xxx.lanl.gov/abs/0812.3622}{{\tt
  arXiv:0812.3622}},
  \href{http://www.slac.stanford.edu/spires/find/hep/www?rawcmd=eprint+0812.36%
22}{[SPIRES]}.

\bibitem{Bezrukov:2008cq}
F.~L. Bezrukov, {\it {The Standard model Higgs as the inflaton}},
  \href{http://xxx.lanl.gov/abs/0805.2236}{{\tt arXiv:0805.2236}},
  \href{http://www.slac.stanford.edu/spires/find/hep/www?rawcmd=eprint+0805.22%
36}{[SPIRES]}.

\bibitem{Shaposhnikov:2008rc}
M.~Shaposhnikov, {\it {Sterile neutrinos in cosmology and how to find them in
  the lab}},  \href{http://dx.doi.org/10.1088/1742-6596/136/2/022045}{{\em J.
  Phys. Conf. Ser.} {\bf 136} (2008)  022045},
  [\href{http://xxx.lanl.gov/abs/0809.2028}{{\tt arXiv:0809.2028}}],
  \href{http://www.slac.stanford.edu/spires/find/hep/www?rawcmd=eprint+0809.20%
28}{[SPIRES]}.

\bibitem{Maiani:1977cg}
L.~Maiani, G.~Parisi, and R.~Petronzio, {\it {Bounds on the Number and Masses
  of Quarks and Leptons}},
  \href{http://dx.doi.org/10.1016/0550-3213(78)90018-4}{{\em Nucl. Phys.} {\bf
  B136} (1978)  115},
  \href{http://www.slac.stanford.edu/spires/find/hep/www?rawcmd=j+NUPHA,B136,1%
15}{[SPIRES]}.

\bibitem{Cabibbo:1979ay}
N.~Cabibbo, L.~Maiani, G.~Parisi, and R.~Petronzio, {\it {Bounds on the
  Fermions and Higgs Boson Masses in Grand Unified Theories}},
  \href{http://dx.doi.org/10.1016/0550-3213(79)90167-6}{{\em Nucl. Phys.} {\bf
  B158} (1979)  295--305},
  \href{http://www.slac.stanford.edu/spires/find/hep/www?rawcmd=j+NUPHA,B158,2%
95}{[SPIRES]}.

\bibitem{Lindner:1985uk}
M.~Lindner, {\it {Implications of Triviality for the Standard Model}},
  \href{http://dx.doi.org/10.1007/BF01479540}{{\em Zeit. Phys.} {\bf C31}
  (1986)  295},
  \href{http://www.slac.stanford.edu/spires/find/hep/www?rawcmd=j+ZEPYA,C31,29%
5}{[SPIRES]}.

\bibitem{Hambye:1996wb}
T.~Hambye and K.~Riesselmann, {\it Matching conditions and higgs mass upper
  bounds revisited},  \href{http://dx.doi.org/10.1103/PhysRevD.55.7255}{{\em
  Phys. Rev.} {\bf D55} (1997)  7255--7262},
  [\href{http://xxx.lanl.gov/abs/hep-ph/9610272}{{\tt hep-ph/9610272}}],
  \href{http://www.slac.stanford.edu/spires/find/hep/www?rawcmd=eprint+hep-ph%
\%2F9610272}{[SPRIES]}.

\bibitem{Krasnikov:1978pu}
N.~V. Krasnikov, {\it {Restriction of the Fermion Mass in Gauge Theories of
  Weak and Electromagnetic Interactions}},  {\em Yad. Fiz.} {\bf 28} (1978)
  549--551,
  \href{http://www.slac.stanford.edu/spires/find/hep/www?rawcmd=j+YAFIA,28,549%
}{[SPIRES]}.

\bibitem{Hung:1979dn}
P.~Q. Hung, {\it {Vacuum Instability and New Constraints on Fermion Masses}},
  \href{http://dx.doi.org/10.1103/PhysRevLett.42.873}{{\em Phys. Rev. Lett.}
  {\bf 42} (1979)  873},
  \href{http://www.slac.stanford.edu/spires/find/hep/www?rawcmd=j+PRLTA,42,873%
}{[SPIRES]}.

\bibitem{Politzer:1978ic}
H.~D. Politzer and S.~Wolfram, {\it {Bounds on Particle Masses in the
  Weinberg-Salam Model}},
  \href{http://dx.doi.org/10.1016/0370-2693(79)90746-9}{{\em Phys. Lett.} {\bf
  B82} (1979)  242--246},
  \href{http://www.slac.stanford.edu/spires/find/hep/www?rawcmd=j+PHLTA,B82,24%
2}{[SPIRES]}.

\bibitem{Altarelli:1994rb}
G.~Altarelli and G.~Isidori, {\it {Lower limit on the Higgs mass in the
  standard model: An Update}},
  \href{http://dx.doi.org/10.1016/0370-2693(94)91458-3}{{\em Phys. Lett.} {\bf
  B337} (1994)  141--144},
  \href{http://www.slac.stanford.edu/spires/find/hep/www?rawcmd=j+PHLTA,B337,1%
41}{[SPIRES]}.

\bibitem{Casas:1994qy}
J.~A. Casas, J.~R. Espinosa, and M.~Quiros, {\it {Improved Higgs mass stability
  bound in the standard model and implications for supersymmetry}},
  \href{http://dx.doi.org/10.1016/0370-2693(94)01404-Z}{{\em Phys. Lett.} {\bf
  B342} (1995)  171--179}, [\href{http://xxx.lanl.gov/abs/hep-ph/9409458}{{\tt
  hep-ph/9409458}}],
  \href{http://www.slac.stanford.edu/spires/find/hep/www?rawcmd=eprint+hep-ph%
\%2F9409458}{[SPRIES]}.

\bibitem{Casas:1996aq}
J.~A. Casas, J.~R. Espinosa, and M.~Quiros, {\it {Standard Model stability
  bounds for new physics within LHC reach}},
  \href{http://dx.doi.org/10.1016/0370-2693(96)00682-X}{{\em Phys. Lett.} {\bf
  B382} (1996)  374--382}, [\href{http://xxx.lanl.gov/abs/hep-ph/9603227}{{\tt
  hep-ph/9603227}}],
  \href{http://www.slac.stanford.edu/spires/find/hep/www?rawcmd=eprint+hep-ph%
\%2F9603227}{[SPRIES]}.

\bibitem{Barvinsky:2008ia}
A.~O. Barvinsky, A.~Y. Kamenshchik, and A.~A. Starobinsky, {\it {Inflation
  scenario via the Standard Model Higgs boson and LHC}},
  \href{http://dx.doi.org/10.1088/1475-7516/2008/11/021}{{\em JCAP} {\bf 0811}
  (2008)  021}, [\href{http://xxx.lanl.gov/abs/0809.2104}{{\tt
  arXiv:0809.2104}}],
  \href{http://www.slac.stanford.edu/spires/find/hep/www?rawcmd=eprint+0809.21%
04}{[SPIRES]}.

\bibitem{GarciaBellido:2008ab}
J.~Garcia-Bellido, D.~G. Figueroa, and J.~Rubio, {\it {Preheating in the
  Standard Model with the Higgs-Inflaton coupled to gravity}},
  \href{http://dx.doi.org/10.1103/PhysRevD.79.063531}{{\em Phys. Rev.} {\bf
  D79} (2009)  063531}, [\href{http://xxx.lanl.gov/abs/0812.4624}{{\tt
  arXiv:0812.4624}}],
  \href{http://www.slac.stanford.edu/spires/find/hep/www?rawcmd=eprint+0812.46%
24}{[SPIRES]}.

\bibitem{Burgess:2009ea}
C.~P. Burgess, H.~M. Lee, and M.~Trott, {\it {Power-counting and the Validity
  of the Classical Approximation During Inflation}},
  \href{http://xxx.lanl.gov/abs/0902.4465}{{\tt arXiv:0902.4465}},
  \href{http://www.slac.stanford.edu/spires/find/hep/www?rawcmd=eprint+0902.44%
65}{[SPIRES]}.

\bibitem{Barbon:2009ya}
J.~L.~F. Barbon and J.~R. Espinosa, {\it {On the Naturalness of Higgs
  Inflation}},  \href{http://xxx.lanl.gov/abs/0903.0355}{{\tt
  arXiv:0903.0355}},
  \href{http://www.slac.stanford.edu/spires/find/hep/www?rawcmd=eprint+0903.03%
55}{[SPIRES]}.

\bibitem{Cornwall:1974km}
J.~M. Cornwall, D.~N. Levin, and G.~Tiktopoulos, {\it {Derivation of Gauge
  Invariance from High-Energy Unitarity Bounds on the s Matrix}},
  \href{http://dx.doi.org/10.1103/PhysRevD.10.1145}{{\em Phys. Rev.} {\bf D10}
  (1974)  1145},
  \href{http://www.slac.stanford.edu/spires/find/hep/www?rawcmd=j+PHRVA,D10,11%
45}{[SPIRES]}.

\bibitem{Longhitano:1980iz}
A.~C. Longhitano, {\it {Heavy Higgs Bosons in the Weinberg-Salam Model}},
  \href{http://dx.doi.org/10.1103/PhysRevD.22.1166}{{\em Phys. Rev.} {\bf D22}
  (1980)  1166},
  \href{http://www.slac.stanford.edu/spires/find/hep/www?rawcmd=j+PHRVA,D22,11%
66}{[SPIRES]}.

\bibitem{Shaposhnikov:2008xi}
M.~Shaposhnikov and D.~Zenhausern, {\it {Quantum scale invariance, cosmological
  constant and hierarchy problem}},
  \href{http://dx.doi.org/10.1016/j.physletb.2008.11.041}{{\em Phys. Lett.}
  {\bf B671} (2009)  162--166}, [\href{http://xxx.lanl.gov/abs/0809.3406}{{\tt
  arXiv:0809.3406}}],
  \href{http://www.slac.stanford.edu/spires/find/hep/www?rawcmd=eprint+0809.34%
06}{[SPIRES]}.

\bibitem{coleman}
S.~Coleman, {\em Dilatations, in Aspects of symmetry}.
\newblock Cambridge University Press, 1985.
\newblock 402 p.

\bibitem{Englert:1976ep}
F.~Englert, C.~Truffin, and R.~Gastmans, {\it {Conformal Invariance in Quantum
  Gravity}},  \href{http://dx.doi.org/10.1016/0550-3213(76)90406-5}{{\em Nucl.
  Phys.} {\bf B117} (1976)  407},
  \href{http://www.slac.stanford.edu/spires/find/hep/www?rawcmd=j+NUPHA,B117,4%
07}{[SPIRES]}.

\bibitem{Jain:2008qv}
P.~Jain, S.~Mitra, and N.~K. Singh, {\it {Cosmological Implications of a Scale
  Invariant Standard Model}},
  \href{http://dx.doi.org/10.1088/1475-7516/2008/03/011}{{\em JCAP} {\bf 0803}
  (2008)  011}, [\href{http://xxx.lanl.gov/abs/0801.2041}{{\tt
  arXiv:0801.2041}}],
  \href{http://www.slac.stanford.edu/spires/find/hep/www?rawcmd=eprint+0801.20%
41}{[SPIRES]}.

\bibitem{Shaposhnikov:2009nk}
M.~E. Shaposhnikov and F.~V. Tkachov, {\it {Quantum scale-invariant models as
  effective field theories}},  \href{http://xxx.lanl.gov/abs/0905.4857}{{\tt
  arXiv:0905.4857}},
  \href{http://www.slac.stanford.edu/spires/find/hep/www?rawcmd=eprint+0905.48%
57}{[SPIRES]}.

\bibitem{Feruglio:1992wf}
F.~Feruglio, {\it {The Chiral approach to the electroweak interactions}},
  \href{http://dx.doi.org/10.1142/S0217751X93001946}{{\em Int. J. Mod. Phys.}
  {\bf A8} (1993)  4937--4972},
  [\href{http://xxx.lanl.gov/abs/hep-ph/9301281}{{\tt hep-ph/9301281}}],
  \href{http://www.slac.stanford.edu/spires/find/hep/www?rawcmd=eprint+hep-ph%
\%2F9301281}{[SPRIES]}.

\bibitem{Coleman:1973jx}
S.~R. Coleman and E.~Weinberg, {\it Radiative corrections as the origin of
  spontaneous symmetry breaking},
  \href{http://dx.doi.org/10.1103/PhysRevD.7.1888}{{\em Phys. Rev.} {\bf D7}
  (1973)  1888--1910},
  \href{http://www.slac.stanford.edu/spires/find/hep/www?rawcmd=j+PHRVA,D7,188%
8}{[SPIRES]}.

\bibitem{Ford1992}
C.~Ford, I.~Jack, and D.~R.~T. Jones, {\it The standard model effective
  potential at two loops},  {\em Nucl.Phys. B} {\bf 387} (1992)
  373--390;Erratum--ibid.B504(1997)551--552,
  [\href{http://xxx.lanl.gov/abs/hep-ph/0111190}{{\tt hep-ph/0111190}}],
  \href{http://www.slac.stanford.edu/spires/find/hep/www?rawcmd=eprint+hep-ph%
\%2F0111190}{[SPRIES]}.

\bibitem{Dutta:2007st}
S.~Dutta, K.~Hagiwara, Q.-S. Yan, and K.~Yoshida, {\it {Constraints on the
  electroweak chiral Lagrangian from the precision data}},
  \href{http://dx.doi.org/10.1016/j.nuclphysb.2007.08.017}{{\em Nucl. Phys.}
  {\bf B790} (2008)  111--137}, [\href{http://xxx.lanl.gov/abs/0705.2277}{{\tt
  arXiv:0705.2277}}],
  \href{http://www.slac.stanford.edu/spires/find/hep/www?rawcmd=eprint+0705.22%
77}{[SPIRES]}.

\bibitem{Kajantie:1995dw}
K.~Kajantie, M.~Laine, K.~Rummukainen, and M.~E. Shaposhnikov, {\it {Generic
  rules for high temperature dimensional reduction and their application to the
  standard model}},  \href{http://dx.doi.org/10.1016/0550-3213(95)00549-8}{{\em
  Nucl. Phys.} {\bf B458} (1996)  90--136},
  [\href{http://xxx.lanl.gov/abs/hep-ph/9508379}{{\tt hep-ph/9508379}}],
  \href{http://www.slac.stanford.edu/spires/find/hep/www?rawcmd=eprint+hep-ph%
\%2F9508379}{[SPRIES]}.

\bibitem{Espinosa:2007qp}
J.~R. Espinosa, G.~F. Giudice, and A.~Riotto, {\it {Cosmological implications
  of the Higgs mass measurement}},
  \href{http://dx.doi.org/10.1088/1475-7516/2008/05/002}{{\em JCAP} {\bf 0805}
  (2008)  002}, [\href{http://xxx.lanl.gov/abs/0710.2484}{{\tt
  arXiv:0710.2484}}],
  \href{http://www.slac.stanford.edu/spires/find/hep/www?rawcmd=eprint+0710.24%
84}{[SPIRES]}.

\bibitem{Luo:2002ey}
M.-x. Luo and Y.~Xiao, {\it {Two-loop renormalization group equations in the
  standard model}},
  \href{http://dx.doi.org/10.1103/PhysRevLett.90.011601}{{\em Phys. Rev. Lett.}
  {\bf 90} (2003)  011601}, [\href{http://xxx.lanl.gov/abs/hep-ph/0207271}{{\tt
  hep-ph/0207271}}],
  \href{http://www.slac.stanford.edu/spires/find/hep/www?rawcmd=eprint+hep-ph%
\%2F0207271}{[SPRIES]}.

\bibitem{Luscher:1987ay}
M.~Luscher and P.~Weisz, {\it {Scaling Laws and Trivality Bounds in the Lattice
  $\phi^4$ Theory. 1. One Component Model in the Symmetric Phase}},
  \href{http://dx.doi.org/10.1016/0550-3213(87)90177-5}{{\em Nucl. Phys.} {\bf
  B290} (1987)  25},
  \href{http://www.slac.stanford.edu/spires/find/hep/www?rawcmd=j+NUPHA,B290,2%
5}{[SPIRES]}.

\bibitem{Amsler:2008zzb}
{\bf Particle Data Group} Collaboration, C.~Amsler {\em et.~al.}, {\it {Review
  of particle physics}},
  \href{http://dx.doi.org/10.1016/j.physletb.2008.07.018}{{\em Phys. Lett.}
  {\bf B667} (2008)  1},
  \href{http://www.slac.stanford.edu/spires/find/hep/www?rawcmd=j+PHLTA,B667,1%
}{[SPIRES]}.

\bibitem{Chetyrkin:1999qi}
K.~G. Chetyrkin and M.~Steinhauser, {\it {The relation between the MS-bar and
  the on-shell quark mass at order alpha(s)**3}},
  \href{http://dx.doi.org/10.1016/S0550-3213(99)00784-1}{{\em Nucl. Phys.} {\bf
  B573} (2000)  617--651}, [\href{http://xxx.lanl.gov/abs/hep-ph/9911434}{{\tt
  hep-ph/9911434}}],
  \href{http://www.slac.stanford.edu/spires/find/hep/www?rawcmd=eprint+hep-ph%
\%2F9911434}{[SPRIES]}.

\bibitem{Melnikov:2000qh}
K.~Melnikov and T.~v. Ritbergen, {\it {The three-loop relation between the
  MS-bar and the pole quark masses}},
  \href{http://dx.doi.org/10.1016/S0370-2693(00)00507-4}{{\em Phys. Lett.} {\bf
  B482} (2000)  99--108}, [\href{http://xxx.lanl.gov/abs/hep-ph/9912391}{{\tt
  hep-ph/9912391}}],
  \href{http://www.slac.stanford.edu/spires/find/hep/www?rawcmd=eprint+hep-ph%
\%2F9912391}{[SPRIES]}.

\bibitem{Smith:1996xz}
M.~C. Smith and S.~S. Willenbrock, {\it {Top quark pole mass}},
  \href{http://dx.doi.org/10.1103/PhysRevLett.79.3825}{{\em Phys. Rev. Lett.}
  {\bf 79} (1997)  3825--3828},
  [\href{http://xxx.lanl.gov/abs/hep-ph/9612329}{{\tt hep-ph/9612329}}],
  \href{http://www.slac.stanford.edu/spires/find/hep/www?rawcmd=eprint+hep-ph%
\%2F9612329}{[SPRIES]}.

\bibitem{Kataev:2009ns}
A.~L. Kataev and V.~T. Kim, {\it {Uncertainties of QCD predictions for Higgs
  boson decay into bottom quarks at NNLO and beyond}},
  \href{http://xxx.lanl.gov/abs/0902.1442}{{\tt arXiv:0902.1442}},
  \href{http://www.slac.stanford.edu/spires/find/hep/www?rawcmd=eprint+0902.14%
42}{[SPIRES]}.

\end{thebibliography}\endgroup

\end{document}